\documentclass[sigplan,10pt]{acmart}
\usepackage{graphicx,mathtools,kantlipsum}
\usepackage[normalem]{ulem}
\usepackage{url}
\usepackage{amsmath,amsthm,amsfonts}
\usepackage{listings}
\usepackage{color}
\usepackage{float}
\usepackage{multirow}
\usepackage{makecell}
\usepackage{algorithm}
\usepackage[noend]{algpseudocode}
\usepackage{todonotes}
\usepackage{subcaption}
\usepackage{booktabs}
\usepackage{caption}
\usepackage[capitalize]{cleveref}
\usepackage{enumitem}
\usepackage{balance}
\usepackage{pifont}
\usepackage[hang,flushmargin]{footmisc} 
\usepackage[toc,page]{appendix}
\renewcommand{\todo}[1]{\textcolor{blue}{TODO: #1}}

\newcommand{\hlc}[1]{\textcolor{black}{#1}}

\newlength{\mysep}
\setlist[enumerate]{noitemsep,topsep=0pt,partopsep=3pt,leftmargin=*}
\setlist[itemize]{noitemsep,topsep=0pt,partopsep=3pt,leftmargin=*}

\begin{document}
\title{Sandslash: A Two-Level Framework for Efficient Graph Pattern Mining}
\author{Xuhao Chen, Roshan Dathathri, Gurbinder Gill, Loc Hoang, Keshav Pingali}
\affiliation{{The University of Texas at Austin} \\
\{cxh,roshan,gill,loc,pingali\}@cs.utexas.edu}
\date{}
\begin{abstract}
Graph pattern mining (GPM) is used in diverse application areas including
social network analysis, bioinformatics, and chemical engineering. Existing GPM
frameworks either provide high-level interfaces for productivity at the cost of
expressiveness or provide low-level interfaces that can express a wide variety
of GPM algorithms at the cost of increased programming complexity.
Moreover, existing systems lack the flexibility to explore combinations of
optimizations to achieve performance competitive with hand-optimized applications.

We present Sandslash, an in-memory Graph Pattern Mining (GPM) framework
that uses a novel programming interface to support productive, expressive, 
and efficient GPM on large graphs. Sandslash provides a high-level API that 
needs only a specification of the GPM problem, and it implements fast 
subgraph enumeration, provides efficient data structures, 
and applies high-level optimizations automatically. 
To achieve performance competitive with expert-optimized implementations, 
Sandslash also provides a low-level API that 
allows users to express algorithm-specific optimizations. This enables 
Sandslash to support both high-productivity and high-efficiency without 
losing expressiveness. We evaluate Sandslash on shared-memory machines 
using five GPM applications and a wide range of large real-world graphs. 
Experimental results demonstrate that applications written using Sandslash 
high-level or low-level API outperform that in state-of-the-art 
GPM systems AutoMine, Pangolin, and Peregrine on average by 13.8$\times$, 
7.9$\times$, and 5.4$\times$, respectively. We also show that 
these Sandslash applications outperform expert-optimized GPM 
implementations by 2.3$\times$ on average with less programming effort.

\end{abstract}

\maketitle
\section{Introduction}
\label{sec:intro}
\begin{figure}[t]
\centering
\includegraphics[width=0.49\textwidth]{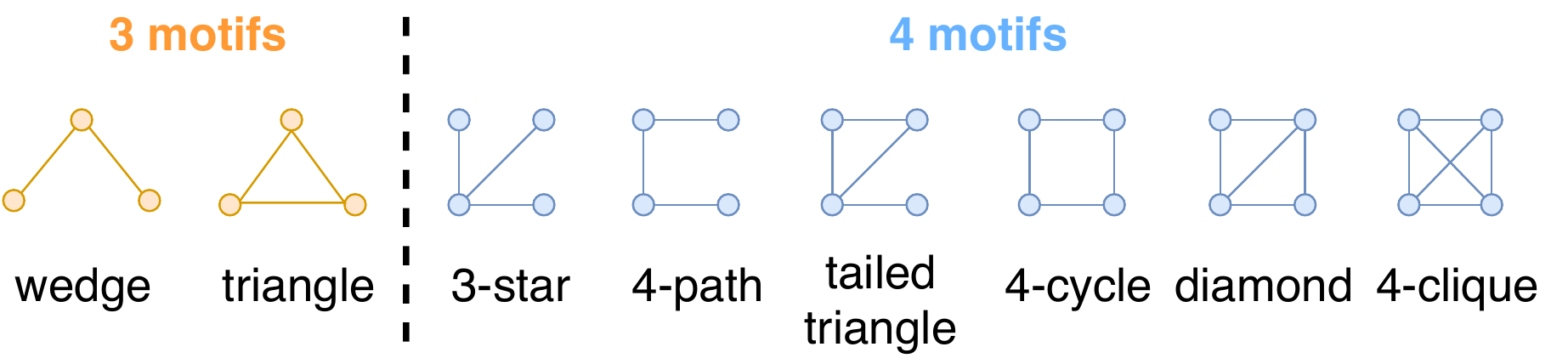}
\caption{3-vertex (left) and 4-vertex (right) motifs.}
\label{fig:motifs}
\end{figure}

Graph pattern mining (GPM) problems arise in many application
domains~\cite{chemical,Motif,Protein,Social}. One example is
{\em motif counting}~\cite{Motifs2,Motif3,Breaking}, which counts
the number of occurrences of certain structural {\em patterns},
such as those shown in \cref{fig:motifs}, in a given graph.
These numbers are often different for graphs from different domains,
so they can be used as a ``signature'' to infer, for example, 
the probable origin of a graph~\cite{Social}.

Programming efficient parallel solutions for GPM problems is challenging.
Most problems are solved by searching the input graph for patterns
of interest. For efficiency, the search space needs to be pruned 
aggressively without compromising correctness, but this can be complicated
since the pruning strategy usually depends on the structural patterns
of interest. Complicated book-keeping data structures are needed
to avoid repeating work during the search process; maintaining these
data structures efficiently in a parallel program can be challenging. 
A number of GPM frameworks have been proposed to reduce the burden on the programmer~\cite{Arabesque,RStream, Kaleido,AutoMine,Pangolin,Peregrine,GraphZero}.
They can be categorized into {\it high-level} and {\it low-level} systems:
they both simplify GPM programming compared to hand-optimized code,
but they make different tradeoffs and have different limitations.

High-level systems such as AutoMine~\cite{AutoMine} and Peregrine~\cite{Peregrine}
take specifications of patterns as input and leverage static analysis
techniques to automatically generate GPM programs for those patterns.
These systems promote productivity, but they may not allow expressing 
more efficient algorithms.
Low-level systems such as RStream~\cite{RStream} and Pangolin~\cite{Pangolin}
provide low-level API functions for the user to control the
details of mining process, and they can be used to implement solutions for
a wider variety of GPM problems but they require more programming effort.
Moreover, both high-level and low-level systems lack the ability to explore
combinations of optimizations that have been implemented in handwritten GPM 
solutions for different problems. Therefore, they do not match the performance 
of hand-optimized solutions.


In this paper, we present \emph{Sandslash}, an in-memory GPM system that provides
high productivity and efficiency without compromising generality. 
Sandslash provides a novel programming interface that separates problem 
specifications from algorithmic optimizations. The Sandslash high-level 
interface requires the user to provide only the specification of the pattern(s) 
of interest. Sandslash analyzes the specification and automatically enables 
efficient search strategies, data representations, and optimizations. To 
customize the GPM algorithm and improve performance further, the user can 
leverage the Sandslash low-level interface to exploit application-specific 
knowledge. This two-level API combines the expressiveness of existing low-level 
systems while achieving the productivity enabled by existing high-level systems.
Meanwhile, Sandslash automates a number of optimizations that have been used 
previously only in handwritten solutions to particular GPM problems.
Since these optimizations are missing in previous high-level systems,
Sandslash can outperform these systems in most cases.
In addition, low-level Sandslash exposes fine-grained control to
allow the user to compose low-level optimizations, which leads to better
performance than hand-optimized implementations without requiring the
programmer to code the entire solution manually.

Evaluation on a 56-core CPU demonstrates that
applications written using Sandslash high-level API
outperform the state-of-the-art GPM systems, AutoMine~\cite{AutoMine},
Pangolin~\cite{Pangolin}, and Peregrine~\cite{Peregrine} by
{7.7$\times$}, 6.2$\times$ and 3.9$\times$ on average, respectively.
Applications using Sandslash low-level API
outperform AutoMine, Pangolin, and Peregrine by
{22.6$\times$}, 27.5$\times$ and 7.4$\times$ on average, respectively.
Sandslash applications are also
2.3$\times$ faster on average than expert-optimized GPM applications.


This work makes the following contributions:
\begin{itemize}

\item We present Sandslash, an in-memory GPM system that supports
productive, expressive, and efficient pattern mining on large graphs.
	
\item We propose a high-level programming model in which the
choice of efficient search strategies, subgraph data structures,
and high-level optimizations are automated, and
we expose a low-level programming model to allow the
programmers to express algorithm-specific optimizations.

\item We holistically analyze the optimization techniques available
in hand-tuned applications and separately enabled them in the
two levels.
\hlc{We show that existing optimizations in the literature
applied to specific problems/applications can be applied
more generally to other GPM problems.}
The user is allowed to flexibly control and explore the
combination of optimizations.

\item Experimental results show that Sandslash substantially
outperforms existing GPM systems.
They also show the impact of (and the need for) Sandslash's low-level API.
Compared to expert-optimized applications, Sandslash achieves
competitive performance with less programming effort.

\end{itemize}

\section{Problem Definition}\label{sect:back}\label{sec:subgraphs}

We use standard definitions in graph theory for {\em subgraph},
{\em vertex-induced} and {\em edge-induced subgraphs},
{\em isomorphism} and {\em automorphism} (definitions
in Appendix \ref{subsect:def}).
Isomorphic graphs $G_1$ and $G_2$ are denoted
as $G_1 \simeq G_2$. If $G_1$ and $G_2$
are automorphic, they are denoted as $G_1 \cong G_2$.

A \emph{pattern} is a graph defined explicitly or
implicitly. An explicit definition specifies the vertices and edges
of the graph while an implicit definition specifies its desired
properties. Given a graph $G$ and a pattern $\mathcal{P}$,
an {\em embedding} $X$ of $\mathcal{P}$ in $G$ is a
vertex- or edge-induced subgraph of $G$
s.t. $X \simeq \mathcal{P}$.
In this work, we focus on connected subgraphs and patterns only.

Given a graph $G$ and a set of patterns $S_p=\{\mathcal{P}_1, \dots,
\mathcal{P}_n\}$, GPM finds all the vertex- or
edge-induced embeddings of $\mathcal{P}_i$ in $G$.
For explicit-pattern problems, the solver finds embeddings of the
given pattern(s). 
For implicit-pattern problems,
$S_p$ is not explicitly given but described using
some rules. Therefore, the solver must find the patterns
as well as the embeddings. If the cardinality of $S_p$ is 1,
we call this problem a {\em single-pattern} problem.
Otherwise, it is a {\em multi-pattern} problem.
Note that \textit{graph pattern matching}~\cite{GPMatch}
finds embeddings only for explicit pattern(s),
whereas \textit{graph pattern mining}~\cite{GPM}
solves both explicit- and implicit-pattern problems.

In some GPM problems, the required output is the listing of embeddings.
However, in other GPM problems, the user wants to get
statistics such as a count of the occurrences of the
pattern(s) in $G$. The particular statistic for $\mathcal{P}$
in $G$ is termed its \textit{support}.
The support has the \emph{anti-monotonic property} if the support of
a supergraph does not exceed the support of a subgraph.
{\em Counting} and {\em listing} may have
different search spaces because listing requires enumerating
every embedding while counting does not.

\begin{figure}[t]
\centering
\includegraphics[width=0.43\textwidth]{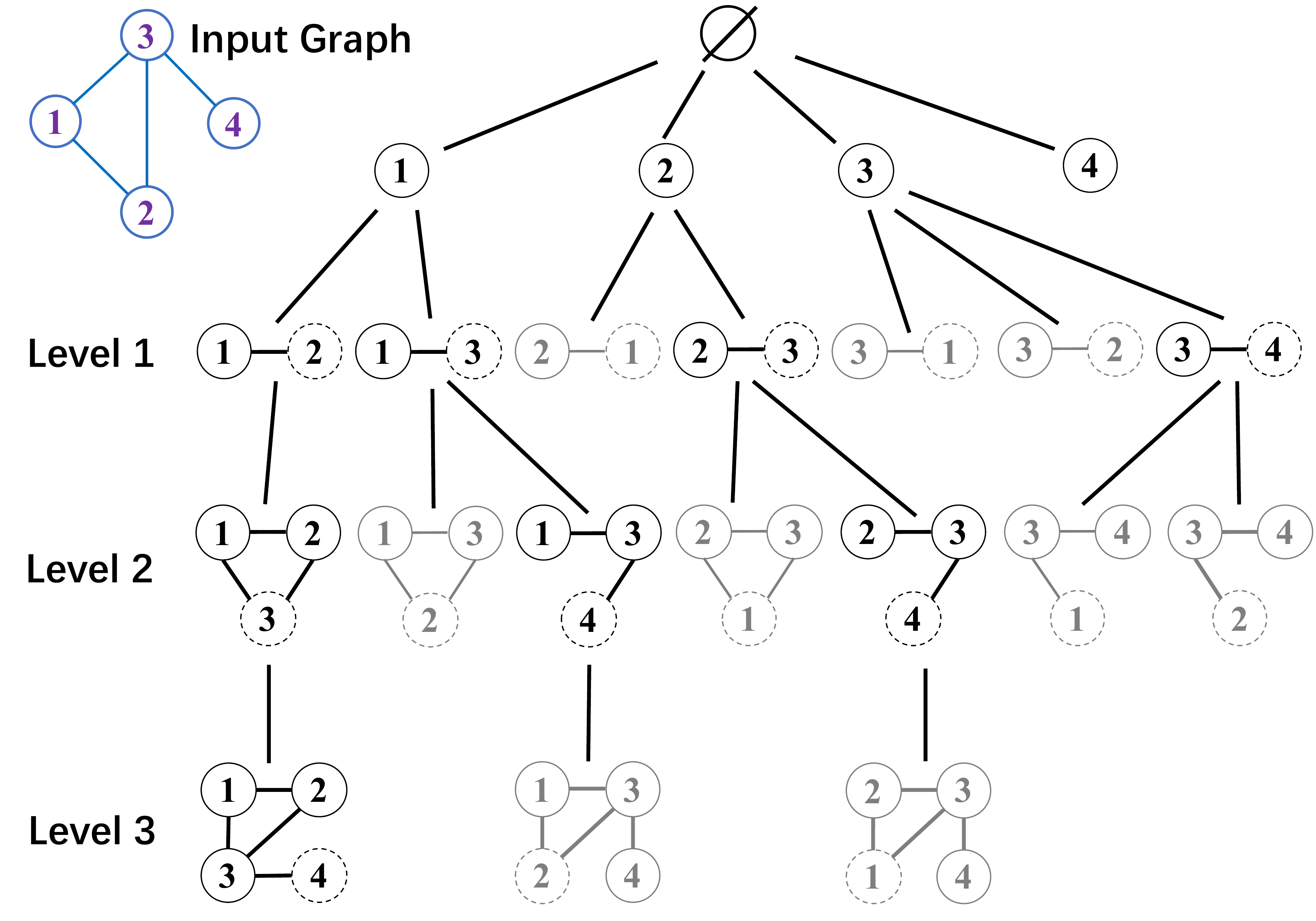}
\caption{\small A portion of a vertex-induced subgraph tree
			with 3 levels. Lightly colored subgraphs are removed
			from consideration by automorphism checks.}
\label{fig:subgraph-tree}
\end{figure}

\paragraph{Graph Pattern Mining Problems}\label{sec:problems}
We consider five GPM problems in a given input graph $G$.

(1) {\em Triangle Counting} (TC):
The problem is to count the number of triangles in $G$.
It uses vertex-induced subgraphs.

(2) {\em $k$-Clique Listing} ($k$-CL):
A subset of vertices $W$ of $G$ is a {\em clique} if every
pair of vertices in $W$ is connected by an edge in $G$. If
the cardinality of $W$ is $k$, this is called a $k$-clique
(triangles are 3-cliques).
The problem of listing $k$-cliques is denoted $k$-CL, and
it uses vertex-induced subgraphs.

(3) {\em Subgraph Listing} (SL):
The problem is to enumerate all {edge-induced} subgraphs of
$G$ isomorphic to a pattern $\mathcal{P}$.

(4) {\em $k$-Motif Counting} ($k$-MC):
This problem counts the number of occurrences of the
different patterns that are possible with $k$ vertices.
In the literature, each pattern is called a {\em motif} or {\em
graphlet}. \cref{fig:motifs} shows all 3-motifs and 4-motifs.
This problem uses vertex-induced subgraphs.

(5) {\em $k$-Frequent Subgraph Mining} ($k$-FSM):
Given integer $k$ and a threshold $\sigma_{min}$ for support,
$k$-FSM finds patterns with $k$ or fewer
edges and lists a pattern $\mathcal{P}$ if its support
is greater than $\sigma_{min}$.
This is called a {\em frequent} pattern.
If $k$ is not specified, it is set to $\infty$, meaning
it considers all possible values of $k$.
This problem finds edge-induced subgraphs.

\subsection{Subgraph Tree and Vertex/Edge Extension}
The \emph{subgraph tree} is a useful abstraction for organizing GPM
computations. The vertex-induced subgraphs of a given input graph $G{=}(V,E)$
can be ordered naturally by containment (\emph{i.e.}, if one
is a subgraph of the other). It is useful to represent this
partial order as a \emph{vertex-induced subgraph tree}
whose vertices represent the subgraphs. Level $l$ of the tree
represents subgraphs with $l+1$ vertices. The root vertex
of the tree represents the empty subgraph\footnote{In practice,
the root and the $0^{th}$ level are not built explicitly.}.
Intuitively, subgraph $S_2{=}(W_2,E_2)$ is a child of subgraph
$S_1{=}(W_1,E_1)$ in this tree if $S_2$ can be obtained by extending
$S_1$ with a single vertex $v \notin W_1$ that is connected to some
vertex in $W_1$ ($v$ is in the {\em neighborhood} of
subgraph $S_1$); this process is called {\em vertex extension}.
Formally, this can be expressed as $W_2 {=} W_1 \cup \{v\}$ where
$v \notin W_1$ and there is an edge $(v,u) \in E$ for some $u \in W_1$.
It is useful to think of the edge connecting $S_1$ and $S_2$ in
the tree as being labeled by $v$. \cref{fig:subgraph-tree} shows
a portion of a vertex-induced subgraph tree with three levels
(for lack of space, not all subgraphs are shown).
Note that a given subgraph can occur in multiple places in this
tree. For example, in \cref{fig:subgraph-tree}, the subgraph
containing vertices 1 and 2 occurs in two places.
These identical subgraphs are automorphisms (automorphic with each other).

The {\em edge-induced subgraph tree} for a given input graph can
be defined in a similar way. \emph{Edge extension} extends an
edge-induced subgraph $S_1$ with a single edge $(u,v)$ provided
at least one of the endpoints of the edge is in $S_1$.

The subgraph tree is a property of the input graph. When solving a specific
GPM problem, a solver uses the subgraph tree as a search tree and builds a
\emph{prefix} of the subgraph tree that depends on the problem, pattern,
and other aspects of the implementation (e.g., if the size of the
pattern is $k$, subgraphs of larger size are not explored).
We use the term {\em embedding tree} to refer to the prefix
of the subgraph tree that has been explored at any point in the search.

Finally, since a pattern is a graph, its connected subgraphs
form a tree as well. These subgraphs are called
{\em sub-patterns}, and the tree formed by them is called the
\emph{sub-pattern tree}.

\definecolor{codegreen}{rgb}{0,0.6,0}
\definecolor{codegray}{rgb}{0.5,0.5,0.5}
\definecolor{codepurple}{rgb}{0.58,0,0.82}
\definecolor{backcolour}{rgb}{1,1,1}
\lstdefinestyle{mystyle}{
	backgroundcolor=\color{backcolour},
	commentstyle=\color{codegreen},
	keywordstyle=\color{blue},
	numberstyle=\tiny\color{codegray},
	stringstyle=\color{codepurple},
	basicstyle=\scriptsize\ttfamily,
	breakatwhitespace=false,
	breaklines=true,
	captionpos=b,
	keepspaces=true,
	numbers=left,
	numbersep=5pt,
	xleftmargin=.05in,
	showspaces=false,
	showstringspaces=false,
	showtabs=false,
	tabsize=2
}
\lstset{style=mystyle}
\lstset{aboveskip=0pt,belowskip=0pt}

\begin{figure}
	\centering
	\includegraphics[width=0.39\textwidth]{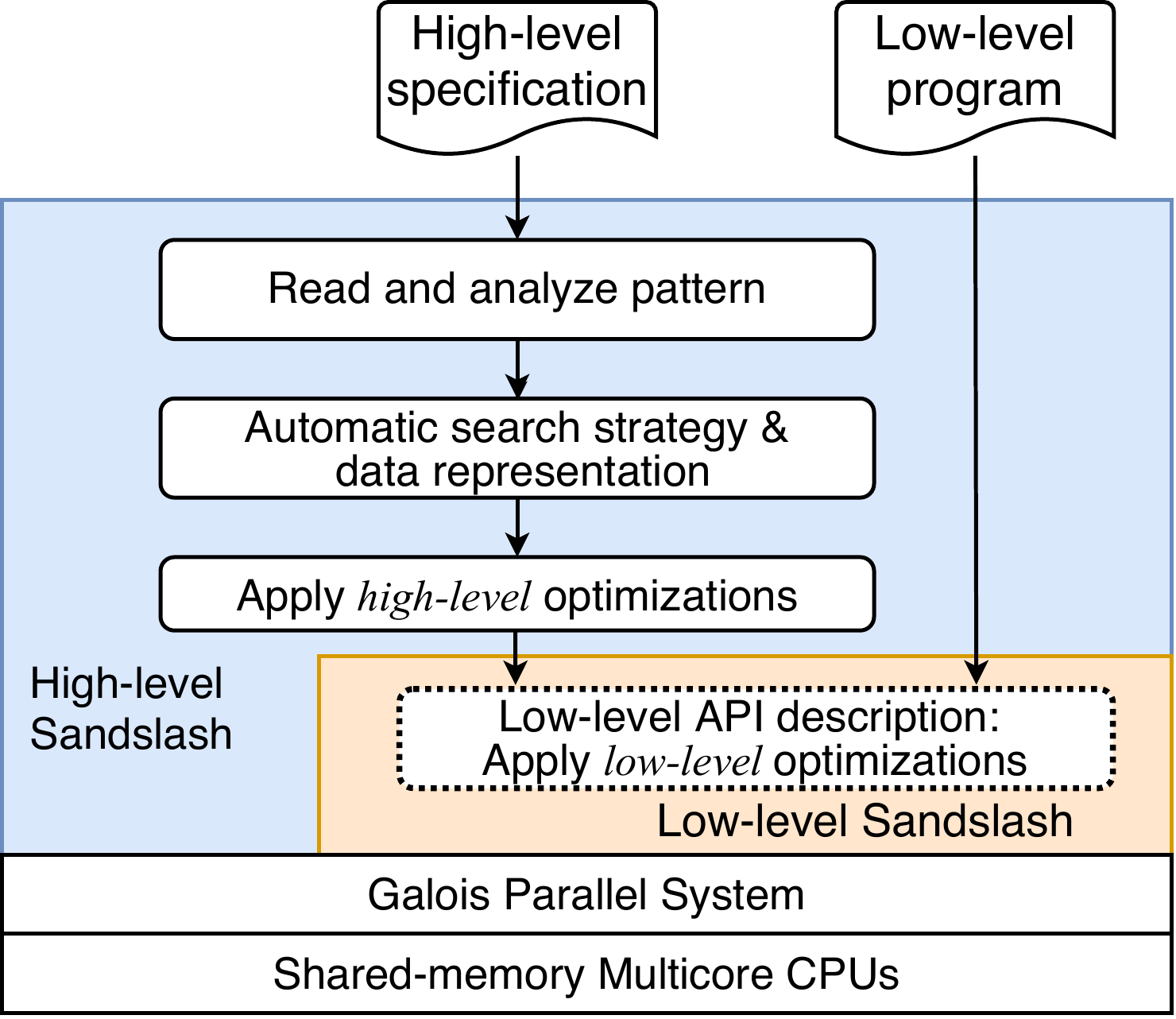}
	\caption{Overview of Sandslash. Dash indicates optional.}
	\label{fig:overview}
\end{figure}

\section{Sandslash API}\label{sec:overview}
\cref{fig:overview} shows the overview of Sandslash, 
which is built on top of the Galois~\cite{Galois} parallel system.
In this section, we describe the high-level and low-level API of Sandslash.

\subsection{High-Level API}\label{sec:hi-api}

\cref{tab:hi-api} shows Sandslash high-level API that 
can be used to specify a GPM problem (as defined in Section~\ref{sect:back}).
The first two required flags define whether the
embeddings are vertex-induced or edge-induced and
whether the matched embeddings need to be listed or counted.
The third required flag defines if the set of patterns is explicit or implicit.
If they are explicit, then the patterns must be defined
using {\tt \small getExplicitPatterns()}.
Otherwise, the rule to select implicit patterns must be defined
using {\tt \small isImplicitPattern()}.
\hlc{{\tt \small process()} is a function for customized output.
{\tt \small terminate()} specifies an optional early termination
condition (useful to implement pattern existence query).
}
The default {\it support} for each pattern in Sandslash is count (number of embeddings).
The {\it support} can be customized using three functions:
{\tt \small getSupport()} defines the support of an embedding,
{\tt \small isSupportAntiMonotonic()} defines if the support
has the anti-monotonic property, and
{\tt \small reduce()} defines the reduction operator (e.g., sum)
for combining the support of different embeddings of the same pattern.
In addition, Sandslash has a runtime parameter $k$
to denote the maximum size (vertices or edges) of the
(vertex- or edge-induced) embeddings to find.

\begin{table}
\centering
\footnotesize
\begin{tabular}{@{}l@{\hspace{5pt}}l@{\hspace{4pt}}||@{\hspace{4pt}}l@{}}
\toprule
{\bf Flag} & {\bf Required} & {\bf Example: spec for FSM} \\
\midrule
isVertexInduced & yes & false \\
isListing & yes & false \\
isExplicit & yes & false \\
\midrule
{\bf Function} & {\bf Required} & {\bf Example: user code for FSM} \\
\midrule
getExplicitPatterns() & no & - \\
isImplicitPattern(Pattern pt) & no & pt.support $\geq$ MIN\_SUPPORT \\
process(Embedding emb) & no & - \\
terminate(Embedding emb) & no & - \\
isSupportAntiMonotonic() & no & true \\
getSupport(Embedding emb) & no & getDomainSupport(emb) \\
reduce(Support s1, Support s2) & no & mergeDomainSupport(s1, s2) \\
\bottomrule
\end{tabular}
\caption{\small Left column lists Sandslash high-level API flags and functions. 
Right column is the spec and user code of FSM using the API.\label{tab:hi-api}}
\end{table}

The problem specifications for TC, CL, SL, and MC are straightforward.
All four specify an explicit set of
patterns\footnote{Sandslash provides helper functions
to enumerate a clique or all patterns of a given size $k$.
It also allows reading the patterns from files.
Sandslash
provides helper functions {\tt getDomainSupport}
and {\tt mergeDomainSupport}
for the standard definition of domain support.}.
Each pattern is specified using an edge-list.
For example, in TC, the user provides an edge-list of \{(0,1) (0,2)
(1,2)\}.
\cref{tab:hi-api}'s right column shows the problem specification for FSM.
{\tt \small isImplicitPattern()} is used to specify that
only {\it frequent} patterns (i.e., those with support greater
than {\tt \small MIN\_SUPPORT}) are of interest.
FSM uses the {\it domain} (MNI) support,
which is anti-monotonic,
and its associated reduce operation.

\subsection{Low-Level API}\label{sec:low-api}

\begin{lstlisting}[float=tp,floatplacement=b,label={lst:lo-api},
language=C++,abovecaptionskip=0pt,belowcaptionskip=0pt,
caption={Sandslash low-level API functions.}]
bool toExtend(Embedding emb, Vertex v);
bool toAdd(Embedding emb, Vertex u);
bool toAdd(Embedding emb, Edge e);
Pattern getPattern(Embedding emb);
void localReduce(int depth, vector<Support> &supports);
void initLG(Graph gg, Vertex v, Graph lg);
void initLG(Graph gg, Edge e, Graph lg);
void updateLG(Graph lg);
\end{lstlisting}

Sandslash low-level API is shown in \cref{lst:lo-api}.
\hlc{Since the mining process includes search, extension,
and reduction, control of the mining process includes
customizing (1) the graph to search ({\tt initLG} and {\tt updateLG}),
(2) the extension candidates and their selection ({\tt toAdd} and {\tt toExtend}),
and (3) the reduction operations to perform ({\tt getPattern} and {\tt localReduce}).
Sandslash low-level API exposes necessary control
and is expressive enough to support sophisticated algorithms.}

{\tt \small toExtend} determines if a vertex $v$ in embedding
$emb$ must be extended. {\tt \small toAdd} decides if extending
embedding $emb$ with vertex $u$ (or edge $e$) is allowed.
{\tt \small toExtend} and {\tt \small toAdd} do fine-grained pruning
to reduce search space (Appendix \ref{subsect:fp}). 

{\hlc{{\tt getPattern()} returns the pattern of an embedding.
This function can be used to replace the default
graph isomorphism test with a custom method
to identify patterns (Appendix \ref{subsect:cp}).
Note that {\tt Pattern} can be user-defined; therefore, Sandslash can
support custom aggregation-keys like Fractal~\cite{Fractal}.}

Some algorithms~\cite{PGD} do local counting for each vertex or edge
instead of global counting. Sandslash provides
{\tt\small localReduce} to support local counting.
Listings~\ref{lst:motif-opt} and \ref{lst:4motif-lc} in the Appendix 
show 3-MC and 4-MC using this low-level API.

Some algorithms~\cite{KClique} search local
(sub-)graphs instead of the (global) input graph.
{\tt initLG} and {\tt updateLG} can be used to
support search on local graphs. \cref{lst:kcl-opt} in the
Appendix shows optimized $k$-CL using the Sandslash low-level API.


\subsection{Productivity and Expressiveness}\label{subsect:loc}

\begin{table}[t]
	\footnotesize
	\centering
		\begin{tabular}{@{}c@{\hskip 1pt}|@{\hskip 1pt}c@{\hskip 1pt}|@{\hskip 1pt}c@{\hskip 3pt}c@{\hskip 3pt}c@{\hskip 3pt}c@{\hskip 3pt}c@{}}
		\toprule
		   & \bf{Language} & \bf{TC}  & \bf{$k$-CL} & \bf{SL} & \bf{$k$-MC} & \bf{$k$-FSM} \\ 
		\midrule
		{\bf Hand} & \multirow{2}{*}{C/C++} & {\scriptsize GAP~\cite{GAPBS}} & {\scriptsize KClist~\cite{KClique}} & {\scriptsize CECI~\cite{CECI}} & {\scriptsize PGD~\cite{PGD}} & {\scriptsize DistGraph~\cite{DistGraph}}  \\
		{\bf Optimized}        &      & 89  & 394 & 3,000 & 2,538 & 17,459   \\\hline
		{\bf Arabesque~\cite{Arabesque}} & Java & 43  & 31  & -     &    35 & 80       \\
		{\bf RStream~\cite{RStream}} & C++  & 101 & 61  & -     &    98 & 125      \\
		{\bf Fractal~\cite{Fractal}} & Java/Scala & 6 & 6 & 5     &    12 & 48       \\
		{\bf Pangolin~\cite{Pangolin}} & C++  & 26  & 36  & 57    &    82 & 252      \\
		{\bf AutoMine~\cite{AutoMine}} & C++  & 0   & 0   & {\tt N/S}     &    -  & {\tt N/S}\\
		{\bf Peregrine~\cite{Peregrine}} & C++  & 0   & 0   & 0     &    54 & 68       \\
		{\bf Sandslash-Hi}& C++& 0   & 0   & 0     &    9  & 75       \\
		{\bf Sandslash-Lo}& C++& -   & 67  & -     &    61 & -      \\
		\bottomrule
		\end{tabular}
	\caption{\small Lines of code in Sandslash, other GPM systems, and 
		expert-optimized GPM applications.
		{\tt N/S}: not supported; `-':  
		supported but not yet implemented.
		AutoMine, Peregrine are high-level systems, Sandslash is a two-level system,
		and the rest are low-level systems.}
	\label{tab:loc}
\end{table}


\cref{tab:loc} compares the lines of code of GPM applications in Sandslash
and other GPM systems with that of hand-optimized GPM applications.
Sandslash's high-level specification requires similar
programming effort to that of other high-level systems. 
Zero lines of C++ user code are needed for explicit pattern 
problems (only flags and pattern edgelist are required).
Sandslash's low-level programs require only a little more programming
effort since all high-level optimizations are
automatically performed by the system.
\hlc{Note that Fractal is classified as a low-level system because it supports
high-level API only for single, explicit-pattern problems (i.e., SL).
For implicit-pattern problems like FSM, Fractal requires
the user to write code for iterative expand-aggregate-filter.}
As for expressiveness, since Sandslash uses the vertex/edge extension
model that existing low-level systems use, it is as expressive as these systems.
Sandslash high-level API is adequate for the specification
of GPM problems implemented in previous GPM systems
and is easily extendable to support new features (e.g., anti-edges and
anti-vertices in Peregrine) that fit in the vertex/edge extension abstraction.
In contrast, existing high-level systems use a pattern-aware
model based on set intersection and set difference
which does not work well for implicit-pattern problems.
For example, FSM implemented in Peregrine and AutoMine
(details in Appendix ~\ref{subsect:mo}) 
needs to enumerate possible patterns before the search, 
which causes non-trivial performance overhead.
In \cref{sec:naive}, we show that the vertex/edge extension model
can be effectively augmented with pattern awareness, and more importantly,
together with the high-level optimizations, Sandslash can even
outperform existing high-level systems, with the same productivity.

\hlc{All existing GPM systems lack support for some of the functions 
in Sandslash low-level API.
Consequently, they either lack such optimizations or expose
an even lower-level API to give the user full control of
the mining process at the cost of preventing the system from
applying high-level optimizations. For example,
Fractal~\cite{Fractal} allows the user to compose low-level optimizations
using custom subgraph enumerators. In contrast, Sandslash
low-level API is designed to enable the user to express custom
algorithms while allowing the system to apply high-level
optimizations and explore different traversal orders.
For example, to implement local graph search in Sandslash,
the user only implements initialization/modification functions
for the local graph, which allows Sandslash to apply all possible
high-level optimizations in \cref{tab:opt} during exploration.
In Fractal, the user must change the entire exploration 
which includes implementing all the optimizations by hand;
the system cannot apply optimizations automatically.
This also applies for other low-level optimizations like local counting. }

\section{High-Level Sandslash}\label{sec:naive}

This section describes high-level Sandslash.
It uses efficient search strategies (\cref{sec:search}),
data representations (\cref{sec:representation}), and
high-level optimizations (\cref{sec:agnostic-opt}).

\subsection{Search Strategies}\label{sec:search}



Given a pattern $\mathcal{P}$ with $k$ vertices, we can build
the subgraph tree described in \cref{sect:back} to a depth $k$ and test each subgraph $X$ at the leaf of the tree to see if $X \simeq \mathcal{P}$.
This approach is pattern-oblivious and works effectively
for any pattern (even if $\mathcal{P}$ is implicit).
In Sandslash, we augment this model with pattern-awareness.
If the user defines an explicit pattern problem,
Sandslash does pattern analysis and constructs a {\em matching
order} (Appendix \ref{subsect:mo}) to guide vertex extension.
This prunes the search tree and avoids isomorphism tests.
As the same subgraph (i.e., automorphisms)
may occur in multiple places of the tree,
Sandslash uses the standard {\em symmetry breaking} technique (see
Appendix \ref{subsect:sym_break}) to avoid over-counting.

GPM solutions must also decide the order in which the subgraph tree is explored.
Any top-down visit order on the tree can be used. Breadth-first
search (BFS) exposes more parallelism while requiring more
storage than a depth-first search (DFS). DFS consumes less
memory but has fewer opportunities for parallel execution.
Arabesque, RStream, and Pangolin use BFS exploration while
the other GPM systems in \cref{tab:loc} use DFS exploration.
Sandslash performs a pseudo-DFS parallel exploration
(following convention this area, we refer to it simply as DFS),
using the following strategy.

\begin{itemize}
\item Each vertex $v$ in the input graph corresponds to a
vertex-induced subgraph for the vertex set $\{v\}$ and
therefore corresponds to a search tree vertex $t_v$.

\item The subtree below each such tree vertex $t_v$ is
explored in DFS order. This is a \emph{task} executed serially
by a single thread. When the exploration reaches the pattern
size $k$, 
the support is updated appropriately.

\item Multiple threads execute different tasks in parallel.
The runtime uses work-stealing for thread load-balancing;
the unit of work-stealing is a task.
\end{itemize}

\paragraph{Pattern filtering for implicit-pattern problems
that use anti-monotonic support:}
The search strategy described above
mines implicit-pattern problems like FSM
by enumerating all embeddings, binning them according
to their patterns, and checking the support for each pattern.
This can be optimized by exploiting the sub-pattern tree
when the support is anti-monotonic (\cref{sec:subgraphs}).
If a sub-pattern does not have enough support, then
its descendants in the sub-pattern tree will not have enough
support and can be ignored due to the anti-monotonic property.
Instead of pruning sub-patterns
during post-processing, one can prune after generating all
the embeddings for a given sub-pattern. This allows Sandslash
to avoid generating the embeddings for descendant sub-patterns.
This is easy in BFS since it generates embeddings level by level,
and in each level, the entire list of the embeddings is scanned
to aggregate support for each sub-pattern.
However, this does not work for DFS of the sub-graph tree as DFS is done by
each thread independently.

To handle pattern-wise aggregation, Sandslash
performs a DFS traversal on the sub-pattern tree instead of the sub-graph tree.
This ensures that
the embeddings for a given sub-pattern are generated by a single
thread using the same approach for {\em pattern extension} in
hand-optimized gSpan~\cite{gSpan}; i.e., the embeddings are gathered to their
pattern bins during extension and canonicality is checked for
each sub-pattern to avoid duplicate pattern enumeration.
When the thread finishes extension in each level, the support
for each sub-pattern can be computed using its own bin of embeddings.

\subsection{Representation of Tree}
\label{sec:representation}

Since subgraphs are created incrementally by vertex or edge
extension of parent subgraphs in the embedding tree, the
representation of subgraphs should allow structure sharing
between parent and child subgraphs. We describe the information
stored in the embedding tree and the
concrete representation of the tree for vertex-induced sub-graphs first.

\begin{figure}[t]
\centering
\includegraphics[width=0.43\textwidth]{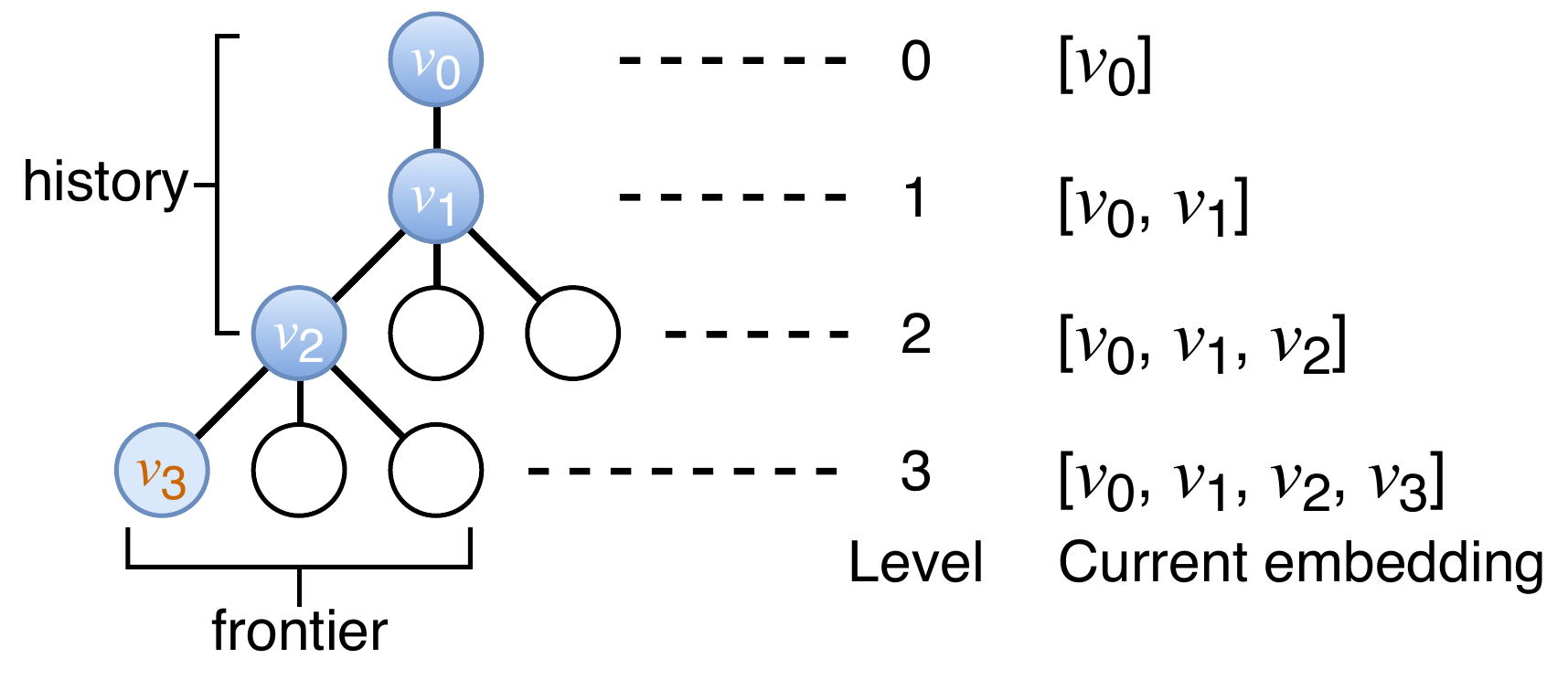}
\caption{Embedding data structure (vertex-induced).}
\label{fig:embedding}
\end{figure}

\begin{itemize}[leftmargin=*]

\item Each non-root vertex in the tree points to its parent vertex.

\item Each non-root vertex in the tree corresponds to
a subgraph obtained from its parent subgraph by vertex extension
with some vertex $v$ of the input graph.
The vertex set of a subgraph can be obtained by walking up the tree and
collecting the vertices stored on the path to the root. These vertices
are the {\em predecessors} of $v$ in the embedding; they correspond
to vertices discovered before $v$ in that embedding.
As shown in \cref{fig:embedding}, the leaf containing
$v_3$ represents the subgraph with a vertex set of $\{v_3, v_2,
v_1, v_0\}$, which are the vertices stored on the path to the
root from this leaf.

\item Given a set of vertices $W=\{v_0, v_1, \dots, v_n\}$ in
a subgraph, the edges among them are obtained from
the input graph $G$. To avoid repetitive look-ups,
edge information is cached  in the embedding tree.
When performing vertex extension by adding a vertex $u$, the edges
between $u$ and its predecessors in the embedding tree are
determined and stored in the tree together with $u$. This set of edges
can be represented compactly using a bit-vector of length $l$ for
vertices at level $l$ of the tree. We call this bit-vector the {\em
connectivity code} (see an example in Appendix \ref{subsect:ccode}).
This technique is called {\em Memoization of Embedding Connectivity} (MEC)~\cite{Pangolin}.

\item For edge-induced extension, a set of
edges instead of vertices is stored for each embedding.
There is no need to store connectivity for embeddings
since the set of edges already contains this information.

\item For a sub-pattern tree, embeddings of each
sub-pattern are gathered as an embedding list (bin of embeddings).
The search tree is constructed with sub-patterns
as vertices, and each sub-pattern in the tree has
an embedding list associated with it. Embedding connectivity
is not needed as the sub-pattern contains this information.

\end{itemize}

\subsection{High Level Optimizations}\label{sec:agnostic-opt}

Sandslash automatically performs high-level optimizations
without guidance from the user. \cref{tab:opt} (left) lists
which of these optimizations are applied to each application.
\cref{tab:sys_opt} (left) lists which of them are
supported by other GPM systems.

\renewcommand\theadalign{bc}
\renewcommand\theadfont{\bfseries}
\renewcommand\theadgape{\Gape[4pt]}
\renewcommand\cellgape{\Gape[4pt]}

\begin{table*}[t]
\footnotesize
\centering
\begin{subtable}[t]{0.37\textwidth}
\centering
		\begin{tabular}{c|c@{ }c@{ }c@{ }c@{ }c|c@{ }c@{ }c@{ }c}
			\Xhline{2\arrayrulewidth}
			        & \multicolumn{5}{c|}{\textbf{High-level}}          & \multicolumn{4}{c}{\textbf{Low-level}} \\ \hline
			        & \bf{SB}    & \bf{DAG}   & \bf{MO}    & \bf{DF}    & \bf{MNC}   & \bf{FP}    & \bf{CP}    & \bf{LG}    & \bf{LC}   \\ \hline
			\textbf{TC}      & \checkmark & \checkmark &            &  \checkmark          &            &            &            &            &           \\
			\textbf{$k$-CL}  & \checkmark & \checkmark & \checkmark & \checkmark & \checkmark &            &            & \checkmark &           \\
			\textbf{SL}      & \checkmark &            & \checkmark & \checkmark & \checkmark &            &            &            &           \\
			\textbf{$k$-MC}  & \checkmark &            &            & \checkmark           & \checkmark & \checkmark & \checkmark &            & \checkmark\\
			\textbf{$k$-FSM} & \checkmark &            &            & \checkmark           &            &            &            &            &           \\
			\Xhline{2\arrayrulewidth}
		\end{tabular}
	\caption{Optimizations applied to GPM applications.}
	\label{tab:opt}
\end{subtable}
\begin{subtable}[t]{0.60\textwidth}
\centering
\begin{tabular}{c|c@{ }c@{ }c@{ }c@{ }c|c@{ }c@{ }c@{ }c}
\Xhline{2\arrayrulewidth}
\multicolumn{1}{l|}{} & \multicolumn{5}{c|}{\textbf{High-level}}                         & \multicolumn{4}{c}{\textbf{Low-level}} \\ \hline
\multicolumn{1}{l|}{} & \textbf{SB}        & \textbf{DAG}       & \textbf{MO}         & \textbf{DF}& \textbf{MNC} & \textbf{FP}          & \textbf{CP}          & \textbf{LG} & \textbf{LC}\\ \hline 
\textbf{AutoMine~\cite{AutoMine}}     &            &            & \checkmark &            & $\circledcirc$ &             &             &             &            \\        
\textbf{Pangolin~\cite{Pangolin}}     & \checkmark & \checkmark & \checkmark &            &  & \checkmark  & \checkmark  &             &            \\        
\textbf{Peregrine~\cite{Peregrine}}    & \checkmark &            & \checkmark &            & $\circledcirc$ &             &             &             &            \\        
\textbf{Sandslash-Hi}        & \checkmark & \checkmark & \checkmark & \checkmark & \checkmark   &             &             &             &            \\        
\textbf{Sandslash-Lo}        & \checkmark & \checkmark & \checkmark & \checkmark & \checkmark   & \checkmark  & \checkmark  & \checkmark  & \checkmark \\        
\Xhline{2\arrayrulewidth}
\end{tabular}
\caption{Optimizations supported by GPM frameworks.  $\checkmark$: full support; $\circledcirc$: limited support.}
\label{tab:sys_opt}
\end{subtable}
\caption{Optimizations enabled in Sandslash.
		\emph{High level optimizations}:
		\textbf{SB}: Symmetry breaking;
		\textbf{DF}: Degree Filtering;
		\textbf{DAG}: orientation;
		\textbf{MO}: Matching Order;
		\textbf{MNC}: Memoization of Neighborhood Connectivity. 
		\emph{Low-level optimizations}:
		\textbf{FP}: Fine-grained Pruning;
		\textbf{CP}: Customized Pattern classification; 
		\textbf{LG}: search on Local Graph; 
		\textbf{LC}: Local Counting.
}
\end{table*}

\noindent{\textbf{Symmetry Breaking (SB), Orientation (DAG), and Matching Order (MO):}}
These optimizations are supported by at least one previous GPM system
(\cref{tab:sys_opt}), so we
describe them in Appendices~\ref{subsect:sym_break}, \ref{subsect:dag},
and~\ref{subsect:mo}.
Sandslash applies SB for all GPM problems.
Sandslash enables DAG only if it is a single explicit-pattern problem
and if the pattern is a clique.
It enables MO for single explicit-pattern problems
except if the pattern is a triangle (because it is not beneficial).

\noindent{\textbf{Degree Filtering (DF):}}
When searching for a pattern in which the
smallest vertex degree is $d$, it is unnecessary to
consider vertices with degree less than $d$ during vertex
extension. When MO is enabled, at each level, only one vertex $v$
of the pattern is searched for, so all vertices with degree less
than that of $v$ can also be filtered. This optimization (DF)
has been used in a hand-optimized SL implementation, PSgL~\cite{PSgL}.
Sandslash enables DF for all GPM problems.
DF is mostly beneficial for SL and $k$-CL with large values of $k$.

\begin{figure}[t]
\centering
\includegraphics[width=0.4\textwidth]{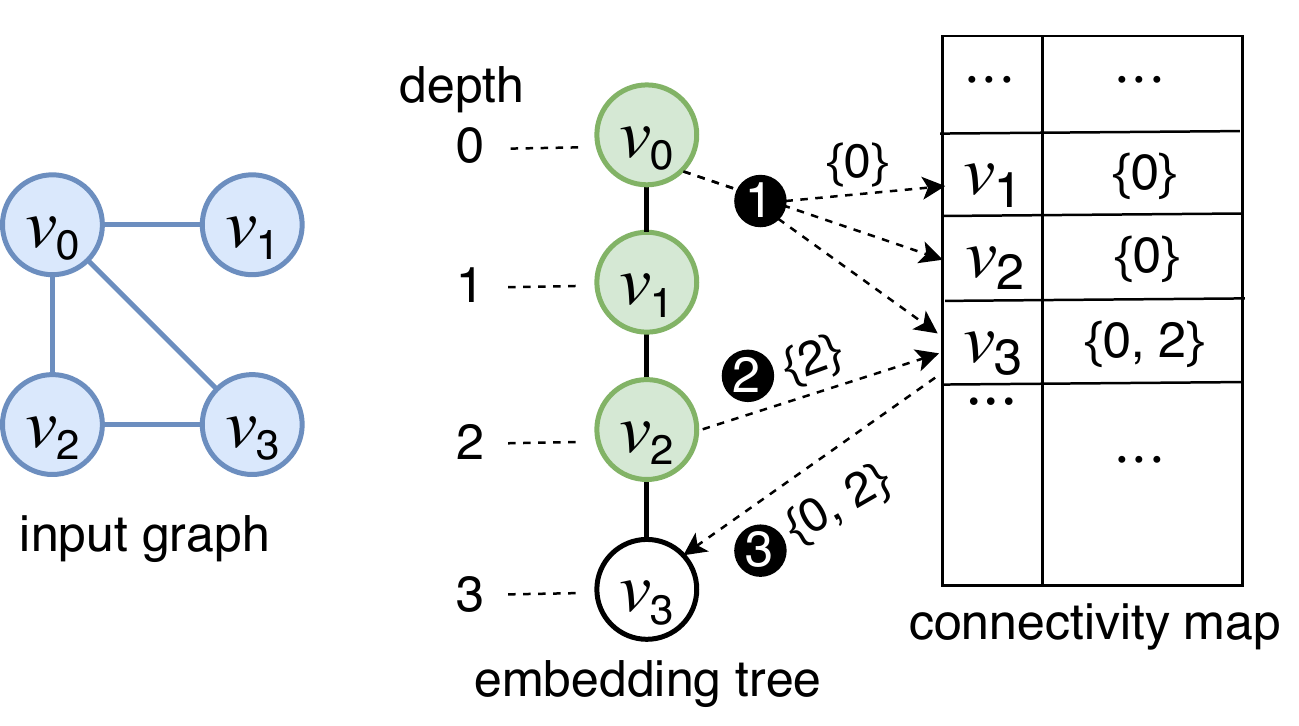}
\caption{An example of connectivity memoization.
			\ding{182}, \ding{183} and \ding{184} are timestamps
			to show the order of actions.}
\label{fig:memo}
\end{figure}

\noindent{\textbf{Memoizing of Neighborhood Connectivity (MNC):}}
When extending an embedding $X=\{v_0, \dots v_n\}$ by adding
a vertex $u$, a common operation
is to check the connectivity between $u$ and each vertex in $X$.
To avoid repeated lookups in the input graph, we memoize
connectivity information in a map (namely {\em connectivity map})
during embedding construction.
The map takes a vertex ID (say $v$) and returns
the positions in the embedding of the vertices connected to $v$.
In \cref{fig:memo}, $v_3$ is connected to $v_0$ and
$v_2$, so when $v_3$ is looked up in the map, the map returns 0
and 2, the embedding positions of $v_0$ and $v_2$.
This map is generated incrementally during the construction of the
embedding tree. Whenever a new vertex ($w$) is added to the
current embedding, the map for the neighbors of $w$ that are not
already in that embedding are updated with the position of $w$
in the embedding; when backing out of this step in the DFS walk,
this information is removed from the map.

\cref{fig:memo} shows how the connectivity map is updated
during vertex extension. At time \ding{182}, depth of $v_0$ is sent to
the map to update the entries of $v_1$, $v_2$ and $v_3$ since $v_1$,
$v_2$ and $v_3$ are neighbors of $v_0$ and they are not in the
current embedding. At time \ding{183}, depth of $v_2$
is sent to the map, and the entry of $v_3$ is updated. Note that
although $v_0$ is also a neighbor of $v_2$, there is no need to update
the entry of $v_0$ since $v_0$ already exists in the current embedding.
When $v_3$ is added to the embedding, the map performs look up
with $v_3$, and the positions \{0, 2\} are returned at time \ding{184}.
Therefore, we know that $v_3$ is connected to the 0-th and 2-th vertices
in the embedding, which are $v_0$ and $v_2$. For parallel execution,
the map is thread private, and each entry is represented by a bit-vector.

This optimization (MNC) has been used in a hand-optimized
$k$-CL implementation, kClist~\cite{KClique}, and a hand-optimized
$k$-MC implementation, PGD~\cite{PGD}.
Sandslash enables this optimization for implicit-pattern problems
that require vertex-induced embeddings and for explicit-pattern
problems unless the pattern is a triangle; for triangles, Sandslash 
uses set intersection instead of MNC. In particular, Sandslash enables MNC for SL, an optimization that is missing in hand-optimized SL implementations~\cite{PSgL,CECI}.

MNC does not exist in previous GPM systems such as Peregrine and
AutoMine which use set intersection/differences to compute connectivity.
Note that MNC is different from the vertex set buffering (VSB) technique
used in Peregrine and AutoMine. To remove redundant computation,
VSB buffers the vertex sets computed for a given embedding.
However, for multi-pattern problems, different patterns may require
buffering different vertex sets. Peregrine's solution is to match one
pattern at a time, which is inefficient for a large number of patterns.
AutoMine's solution is to only buffer one vertex set,
which leads to recomputing unbuffered vertex sets.
The other alternative is to buffer multiple vertex sets for a large pattern,
but this does not scale in terms of memory use.
Unlike these solutions, MNC works well for multi-pattern problems
since the information in the map can be used for both set
intersection and set difference. 

\section{Low-Level Sandslash}\label{sect:lo}

Hand-optimized GPM applications~\cite{KClique,Pivoting,PGD,ESCAPE} can 
use algorithmic insight to aggressively prune the search tree.
\cref{tab:opt} (right) lists optimizations that are applicable to each
application, and \cref{tab:sys_opt} (right) lists which of them are 
supported by other GPM systems. 
In this section, we describe how 
Sandslash low-level API enables users to express such 
optimizations without implementing everything from scratch. 
Sandslash low-level API also
allows Sandslash to perform all possible high-level optimizations, 
which leads to even better performance than hand-optimized applications. 
To use the low-level API, the user needs only to understand 
the subgraph tree abstraction and how to prune the tree. 
They do not need to understand Sandslash's implementation.




\noindent{\textbf{Fine-Grained Pruning (FP) and 
		Customized Pattern Classification (CP):}}
FP and CP are low-level optimizations enabled in a prior 
system, Pangolin~\cite{Pangolin}, so we describe them in
Appendices~\ref{subsect:fp} and~\ref{subsect:cp}. To support FP, Sandslash
exposes API calls {\tt \small toExtend()} and {\tt \small toAdd()} (\cref{lst:lo-api}), 
which allow the user to use algorithmic insight to prune the search space. 
FP is enabled when either function is used. To support CP, Sandslash 
exposes {\tt \small getPattern()} to identify (or classify) the pattern of an
embedding using pattern features instead of  
graph isomorphism tests. CP is enabled when this function is used.

\begin{figure}[t]
\centering
\includegraphics[width=0.23\textwidth]{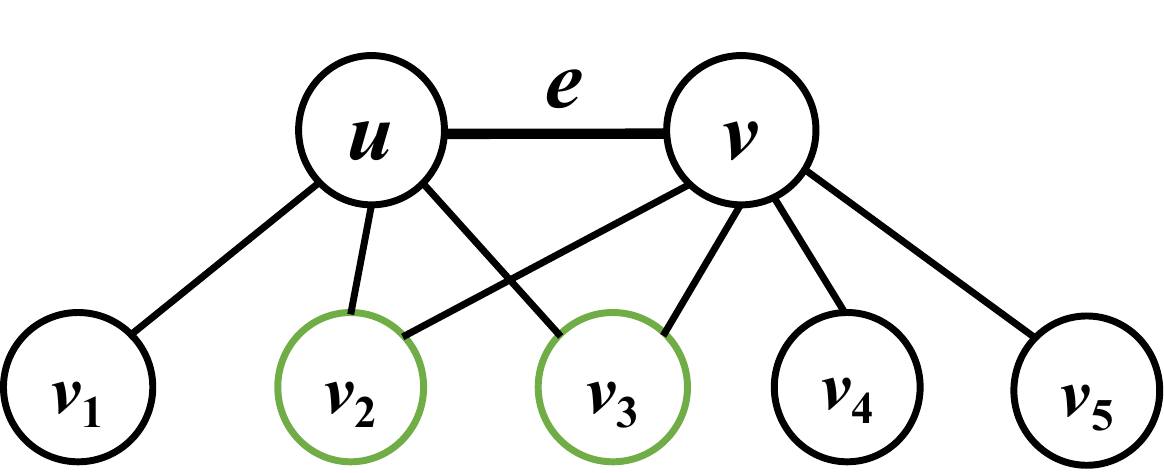}
\caption{\small An example of local counting.
			Given the local triangle count of edge $e$ is $C_{tri}=2$,
			and $deg_{u}=4$, $deg_{v}=5$, we can get local wedge count of $e$ as 
			$C_{wdg}=(4-2-1)+(5-2-1)=3$.}
\label{fig:formula-example}
\end{figure}

\noindent{\textbf{Local Counting (LC):}}
For GPM problems that count matched embeddings instead of listing them,
there may be no need to enumerate all matched embeddings since it 
may be possible to derive precise counts from counts of other patterns.
Formally, the count of embeddings that match a pattern $\mathcal{P}$ may 
be calculated using the count of embeddings that match another pattern
$\mathcal{P'}$. This is useful when both patterns are being searched 
for or when one pattern is more efficient to search for than the other.
This typically requires a {\em local count}~\cite{Pivoting}
(or {\em micro-level count}~\cite{PGD}) of embeddings
associated with a single vertex or edge
instead of a {\em global count} (or {\em macro-level count})
of embeddings that match the pattern.


\setlength{\abovedisplayskip}{3.5pt}
\setlength{\belowdisplayskip}{3.5pt}

Given a pattern $\mathcal{P}$ and a vertex $v$ (or an edge $e$) $\in G$, 
let $S$ be the set of all the embeddings of $\mathcal{P}$ in $G$. 
The {\em local count} of $\mathcal{P}$ on $v$ (or $e$) is defined as 
the number of subgraphs in $S$ that contains $v$ (or $e$).
\cref{fig:formula-example} shows an example of local counting on edge $e$.
Given an edge $e:(u, v)$, the local count of $e$ for
wedges $C_{wdg}(e)$ can be calculated from the local count
of $e$ for triangles $C_{tri}(e)$ using this formula:
\begin{equation}\label{eq:wdg}
C_{wdg} = (deg_{u} - C_{tri} - 1) + (deg_{v} - C_{tri} - 1)
\end{equation}
$deg_{u}$ and $deg_{v}$ are the degrees of $u$ and $v$.

Since wedge counts can be computed from triangle counts,
enumerating wedges is avoided when using local counting for 3-MC.
Similar formulas can be applied for $k$-MC.
Besides, local counting can also be used to prune the 
search space for many subgraph counting problems. For example,
to count edge-induced diamonds, we first compute 
the local triangle count $n_t$ for each edge $e$, and use the formula 
$n_t \choose 2$ $= n_t \times (n_t-1) / 2$ to get the local diamond count. 
The global diamond count is obtained by simply accumulating local counts.


Sandslash exposes {\tt\small localReduce} (\cref{lst:lo-api}) to 
let the user specify how local counts are accumulated.
Sandslash also exposes {\tt \small toExtend} and {\tt \small toAdd} 
to permit the user to customize the subgraph tree exploration
so that the user can determine which patterns need to be enumerated.
\cref{lst:motif-opt} in Appendix shows the user code for 3-MC 
using local counting. Sandslash activates local counting when the user 
implements {\tt\small localReduce}.

\begin{figure}[t]
\centering
\includegraphics[width=0.4\textwidth]{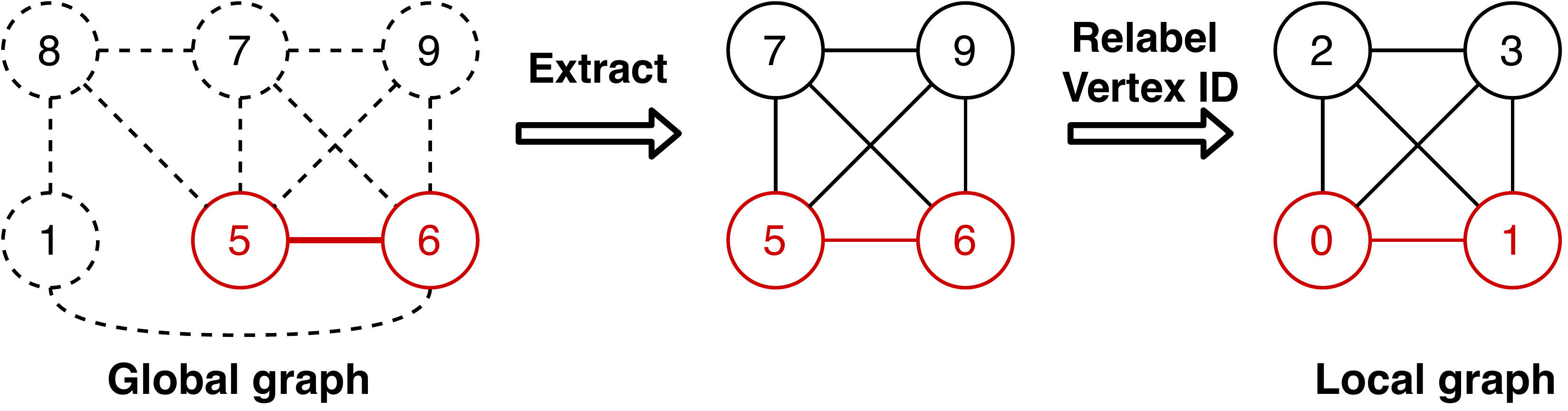}
\caption{\small Local graph induced by edge $(v_5,v_6)$
			and common neighbors of $v_5$ and $v_6$ from the
			global graph.}
\label{fig:local-graph}
\end{figure}

\noindent{\textbf{Search on Local Graph (LG)}}\label{subsect:local-graph}
In some problems like $k$-CL, pattern invariants can be exploited to prune 
the search space~\cite{KClique}. To extend an embedding \{$v_1...v_i$\} 
at level $i-1$ for $k$-CL, the baseline search strategy considers all the neighbors 
of these vertices in the input graph that are not already in the embedding. 
For $k$-CL, any successful candidate vertex for extension must be a neighbor 
of all $v_i$ (otherwise the extension will not result in a clique), so it is 
more efficient to intersect the neighbor lists of the $v_i$ and consider 
only the vertices in this intersection as candidates for vertex extension. 
Abstractly, this can be viewed as constructing a {\em local graph} that is the 
vertex-induced subgraph $G_i$ consisting of the vertices $v_i$ and the vertices 
in the intersection of their neighbor lists, and selecting candidate vertices 
only from $G_i$. Furthermore, if $v_{i+1}$ is the vertex selected at this stage, 
subgraph $G_{i+1}$ is obtained from $G_i$ by removing all vertices in $G_i$ 
that are not neighbors of $v_{i+1}$. In this way, the local graph keeps 
shrinking as the level increases, further reducing the search space.
\cref{fig:local-graph} illustrates an example of induced local graphs.

Sandslash allows the user specify how to initialize
the local graph using {\tt \small initLG()} and update
it at the end of each DFS level using {\tt \small updateLG()} (optional).
When {\tt \small initLG()} is defined, Sandslash enables LG to get the 
neighborhood information during extension using the local graph. 

\begin{table}[t]
	
	\footnotesize
	\centering
		\begin{tabular}{crrrrr}
			\toprule
			\bf{Graph} & \bf{Source} & \multicolumn{1}{c}{\bf{\# V}} & \multicolumn{1}{c}{\bf{\# E}} & \multicolumn{1}{c}{\bf{$\overline{d}$}} & \bf{\# Labels}\\
			\midrule
			\texttt{Pa} & {Patents}~\cite{Patent} & 3M & 28M & 10 & 37\\
			\texttt{Yo} & {Youtube}~\cite{Youtube} & 7M & 114M & 16 & 29\\
			\texttt{Pdb}& {ProteinDB}~\cite{DistGraph}  & 49M & 388M & 8 & 25\\
			\texttt{Lj} & {LiveJournal}~\cite{SNAP} & 5M & 86M & 18 & 0\\
			\texttt{Or} & {Orkut}~\cite{SNAP} & 3M & 234M & 76 & 0\\
			\texttt{Tw4} & {Twitter40}~\cite{twitter40} & 42M	& 2,405M & 29 & 0\\
			\texttt{Fr} & {Friendster}~\cite{friendster} & 66M & 3,612M & 28 & 0\\
			\texttt{Uk} & {UK2007}~\cite{uk2007} & 106M & 6,604M & 31 & 0\\
			\texttt{Gsh}& {Gsh-2015}~\cite{gsh2015} & 988M & 51,381M & 52 & 0\\
			\bottomrule
		\end{tabular}
	\caption{\small Input graphs (symmetric, no loops, 
		no duplicate edges, neighbor list sorted) and their
		properties ($\overline{d}$ is the average degree).}
	\label{tab:input}
\end{table}

\section{Evaluation}\label{sect:eval}

We present experimental setup in \cref{sec:exp-setup}
and then compare Sandslash with state-of-the-art GPM systems
and expert-optimized implementations in \cref{sec:third-party}.
Finally, Sandslash is analyzed in more detail in Sections~\ref{subsect:eval-opt}.

\subsection{Experimental Setup}
\label{sec:exp-setup}

We evaluate two variants of Sandslash:
Sandslash-Hi, which only enables high-level optimizations, and
Sandslash-Lo, which enables both high-level and low-level optimizations.
We compare Sandslash with the state-of-the-art GPM
systems\footnote{These GPM systems are orders of magnitude faster 
than previous GPM systems such as Arabesque \cite{Arabesque}, 
RStream~\cite{RStream}, G-Miner~\cite{G-Miner}, and Fractal~\cite{Fractal}.}:
AutoMine~\cite{AutoMine}, Pangolin~\cite{Pangolin}, 
and Peregrine~\cite{Peregrine}. 
We use the five applications (also used in previous systems) listed in \cref{tab:loc}.
We also evaluate the state-of-the-art expert-optimized GPM 
applications~\cite{GAPBS,KClique,PGD,DistGraph} listed in \cref{tab:loc}
except for CECI~\cite{CECI} (not publicly available).
For fair comparison, we modified DistGraph~\cite{DistGraph} 
and PGD~\cite{PGD} so that 
they produce the same output as Sandslash. We added a parameter $k$ 
in DistGraph: exploration stops when the pattern size becomes $k$. 
For PGD, we disabled counting disconnected patterns.



\cref{tab:input} lists the input graphs.
The first 3 graphs ({\tt Pa}, {\tt Yo}, {\tt pdb})
are vertex-labeled graphs which can be used for FSM.
We also include widely used large graphs ({\tt Lj}, {\tt Or},
{\tt Tw4}, {\tt Fr}, {\tt Uk}), and a very large
web-crawl~\cite{gsh2015} (\texttt{Gsh}). These graphs
do not have labels and are only used for TC, $k$-CL, SL, $k$-MC.

Our experiments were conducted on a 4 socket machine with Intel Xeon Gold
5120 2.2GHz CPUs (56 cores in total) and 190GB RAM. 
All runs use 56 threads.
For the largest graph, \texttt{Gsh}, we used a 2 socket machine with Intel
Xeon Cascade Lake 2.2 Ghz CPUs (48 cores in total)
and 6TB of Intel Optane PMM (byte-addressable memory technology).

Peregrine preprocesses the input graph to reorder vertices 
based on their degrees, which can improve the performance of 
GPM applications. In our evaluation, 
Sandslash does not reorder vertices to be fair to other systems 
and hand-optimized applications which do not perform such preprocessing. 
We use a time-out of 30 hours,
exclude graph loading and preprocessing 
time,
and report results as an average of three runs.


\subsection{Comparisons with Existing Systems}
\label{sec:third-party}


Recall that Tables~\ref{tab:opt} and~\ref{tab:sys_opt} list the optimizations
applicable for each GPM application and enabled by each GPM system.

\noindent{\textbf{Triangle Counting (TC):}}
Note that BFS and DFS are similar for enumerating triangles.
As shown in \cref{tab:tc}, Sandslash achieves competitive performance 
with Pangolin and expert-implemented GAP~\cite{GAPBS}.
Both Pangolin and Sandslash outperform Peregrine and AutoMine 
because they use DAG which is more efficient than on-the-fly symmetry breaking.
On average, Sandslash outperforms AutoMine, Pangolin, Peregrine, and GAP 
by 10.1$\times$, 1.4$\times$,  13.8$\times$, and 1.4$\times$ respectively, for TC.

\begin{table}[t]
	\footnotesize
	\centering
		\begin{tabular}{c|rrrrr}
			\toprule
										& \textbf{Lj}   & \textbf{Or}   
										& \textbf{Tw4}   & \textbf{Fr}   & \textbf{Uk}   \\ \midrule
			\textbf{Pangolin}           & 0.4 & 2.3 
			& 75.5 & 55.1 & 45.8 \\
			\textbf{AutoMine}          & 1.1 & 6.4 
			& 9849.4 & 126.6 & 565.9 \\
			\textbf{Peregrine}          & 1.6 & 7.3 
			& 8492.4 & 100.3 & 3640.9 \\
			\textbf{GAP}   & 0.3 & 2.7 
			& 65.8 & 77.0 & 48.1 \\
			\textbf{Sandslash-Hi}       & \textbf{0.3} & \textbf{1.8} 
			& \textbf{57.2} & \textbf{44.9} & \textbf{24.5} \\
			\bottomrule
		\end{tabular}
	\caption{\small Execution time (sec) of TC.}
	\label{tab:tc}
\end{table}

\begin{table}[t]
	\footnotesize
	\centering
		\begin{tabular}{@{\hskip 3pt}c@{\hskip 3pt}|@{\hskip 3pt}r@{\hskip 3pt}r@{\hskip 3pt}r@{\hskip 3pt}r@{\hskip 3pt}r@{\hskip 3pt}|@{\hskip 3pt}r@{\hskip 3pt}r@{\hskip 3pt}r@{\hskip 3pt}}
		\toprule
		                      & \multicolumn{5}{c|@{\hskip 3pt}}{\textbf{4-CL}}        & \multicolumn{3}{c}{\textbf{5-CL}}         \\ \cline{2-9}
						 & \textbf{Lj} & \textbf{Or} 
						 & \textbf{Tw4} & \textbf{Fr} & \textbf{Uk} & \textbf{Lj} & \textbf{Or}  & \textbf{Fr} \\ \midrule 
		\textbf{Pangolin}     & 19.5      & 56.6        
		& TO       & 564.1     & TO        & 970.4     &  223.4     & 1704.4    \\          
		\textbf{AutoMine}    & 11.0      & 32.9        
		& 67168.4   &  209.6 & 44666.6 & 575.6       &  170.1         &   389.0      \\
		\textbf{Peregrine}    & 15.9      & 73.7        
		& TO       & 397.3     & 55808.4 & 520.8     & 782.1      & 957.6     \\          
		\textbf{kClist}       & 1.2       & 2.5         
		& 1174.0   & 84.0      & OOM         & 22.3      & 5.8        & 87.5      \\          
		\textbf{Sandslash-Hi} & \textbf{0.6} & 2.4 
		& 1676.8        & {166.2}   & 2481.2    & \textbf{13.9} & {7.4}  & {194.9}   \\          
		\textbf{Sandslash-Lo} & 0.7 & \textbf{1.9} 
		& \textbf{681.8} & \textbf{60.4} & \textbf{2451.7} & {14.2} & \textbf{4.8} & \textbf{64.3} \\      
		\bottomrule
		\end{tabular}%
	\caption{\small $k$-CL execution time (sec)
		({\footnotesize OOM: out of memory; TO: timed out}).}
	\label{tab:cl}
\end{table}

\noindent{\textbf{$k$-Clique Listing ($k$-CL):}}
\cref{tab:cl} presents $k$-CL results.
Pangolin (BFS-only) performs poorly as memoizing of neighborhood 
connectivity (MNC) can only be enabled in DFS. Peregrine does 
on-the-fly symmetry breaking (SB), but it does not construct and use 
DAG as in Pangolin and Sandslash. Therefore, Peregrine performs 
similarly to Pangolin although it is a DFS based system. 
AutoMine is slower than
Sandslash because it does not do symmetry breaking. We observe that 
Sandslash-Hi is already significantly faster than all the previous GPM systems. 
Moreover, Sandslash-Lo  
achieves better performance than even expert-implemented kClist~\cite{KClique}
by enabling search on a local graph (LG).
There are some cases where Sandslash-Lo underperforms Sandslash-Hi.
This is because searching on local graph requires computing/maintaining 
local graphs. When the search space is not reduced significantly,
the overhead might outweigh the benefits. We explain this in detail in 
\cref{subsect:eval-opt}. On average, Sandslash-Lo outperforms AutoMine, 
Pangolin, Peregrine, and kClist by 21.0$\times$, 35.1$\times$, 
31.1$\times$, and 1.4$\times$ respectively, for $k$-CL.

\begin{table}[t]
	\footnotesize
	\centering
	\begin{tabular}{@{\hskip 3pt}c@{\hskip 3pt}|@{\hskip 3pt}r@{\hskip 5pt}r@{\hskip 5pt}r@{\hskip 5pt}r@{\hskip 5pt}r@{\hskip 3pt}|@{\hskip 3pt}r@{\hskip 5pt}r@{\hskip 3pt}}
	\toprule
	          & \multicolumn{5}{c|@{\hskip 3pt}}{\bf 3-MC}& \multicolumn{2}{c}{\bf 4-MC} \\ \cline{2-8}
				  & {\bf Lj} & {\bf Or} 
				  & {\bf Tw4} & {\bf Fr} & {\bf Uk} & {\bf Lj} & {\bf Or} \\ \midrule
	{\bf Pangolin}      & 10.8        & 96.5          
	& TO               & 2460.1         & 23676.6        & TO            & TO             \\
	{\bf AutoMine}     & 3.1					 &  18.2      
	& 48901.7 &  352.8        &   4051.0     &   15529.7            & 90914.5             \\
	{\bf Peregrine}     & 2.5         & 4.9  
	& 8447.4           & 165.3          & 3571.5             & 163.6         & 1701.4         \\
	{\bf PGD}           & 11.2        & 42.5          
	& OOM                & OOM              & OOM              & 192.8         & 4069.6         \\
	{\bf Sandslash-Hi}  & 2.1         & 12.1          
	& TO               & 723.7          & 4979.1         & 2366.2        & 30394.7        \\
	{\bf Sandslash-Lo}  & \textbf{0.3} & \textbf{1.6} 
	& \textbf{304.6} & \textbf{43.8} & \textbf{386.8} & \textbf{16.7} & \textbf{232.4}\\
	\bottomrule
	\end{tabular}
\caption{\small $k$-MC execution time (sec)
({\footnotesize OOM: out of memory; TO: timed out}).}
\label{tab:mc}
\end{table}

\begin{table}[t]
	\footnotesize
	\centering
		\begin{tabular}{@{\hskip 3pt}c@{\hskip 3pt}|@{\hskip 3pt}r@{\hskip 5pt}r@{\hskip 5pt}r@{\hskip 5pt}r@{\hskip 3pt}|@{\hskip 3pt}r@{\hskip 5pt}r@{\hskip 5pt}r@{\hskip 3pt}}
			\toprule
			& \multicolumn{4}{c|@{\hskip 3pt}}{\textbf{diamond}}  & \multicolumn{3}{c}{\textbf{4-cycle}}                                                                                     \\ \cline{2-8}
			& \textbf{Lj} & \textbf{Or} 
			& \textbf{Tw4} & \textbf{Fr} & \textbf{Lj} & \textbf{Or} & \textbf{Fr} \\ \hline
			\textbf{Pangolin}  & 92.3 & 884.5 
			& TO & 9301.6 & 553.5 & 13208.2 & TO \\
			\textbf{Peregrine} & 5.4  & 10.2  
			& \textbf{20898.4} & \textbf{178.1} & 144.4 & 1867.2 & 32276.8 \\
			\textbf{Sandslash-Hi} & \textbf{1.5} & \textbf{4.2} 
			& 44659.5 & 284.2 & \textbf{6.3} & \textbf{79.0} & \textbf{20490.9} \\ 
			\bottomrule
		\end{tabular}
	\caption{\small Execution time (sec) of SL.}
	\label{tab:sl}
\end{table}

\begin{table*}[t]
\footnotesize
\centering
\begin{tabular}{@{}c@{\hskip 3pt}|rrr|rrr|rrr|rrr|rrr@{}}
\toprule
\multicolumn{1}{l}{} & \multicolumn{9}{|c|}{\textbf{3-FSM}}                & \multicolumn{6}{c}{\textbf{4-FSM}} \\ 
\cline{2-16}
\multicolumn{1}{l}{} 
& \multicolumn{3}{|c|}{\textbf{Pa}}                & \multicolumn{3}{c|}{\textbf{Yo}} & \multicolumn{3}{c|}{\textbf{Pdb}}   & \multicolumn{3}{c|}{\textbf{Pa}}                  & \multicolumn{3}{c}{\textbf{Pdb}}                    \\ \hline
\textbf{$\sigma_{min}$}            
& \textbf{500} & \textbf{1K} & \textbf{5K} & \textbf{500} & \textbf{1K} & \textbf{5K} & \textbf{500} & \textbf{1K} & \textbf{5K} & \textbf{10K}  & \textbf{20K} & \textbf{30K} & \textbf{500} & \textbf{1K} & \textbf{5K} \\ \midrule
\textbf{Pangolin}                  
& 17.0                 & 19.1                  & 27.4                           & 86.8         & 88.3        & 91.5    & 57.6&66.1&117.3    & OOM                   & 146.2        & 29.4         & OOM           & OOM                     & OOM           \\
\textbf{Peregrine}                  
& 103.8               & 118.4                  & 94.3                           & \textbf{52.8}         & \textbf{69.9}       & \textbf{60.8} &928.7&837.1&943.7        & 28301.0         & 4240.6       & 397.3                 & TO           & TO          & TO          \\
\textbf{DistGraph}                  
& 13.1                 & 13.0                  & 14.1                              & OOM            & OOM           & OOM     &253.9&278.8&239.8      & 120.7                & \textbf{58.01}        & \textbf{25.1}                   & OOM            & OOM           & OOM          \\ 
\textbf{Sandslash}                  
& \textbf{3.5}          & \textbf{3.8}         & \textbf{6.1}                  & 81.0         & 80.8        & 82.8      &\textbf{46.5}&\textbf{40.0}&\textbf{44.5}  & \textbf{102.3}               & 108.4        & 43.7            & \textbf{200.2}        & \textbf{198.0}       & \textbf{195.1}       \\
\bottomrule
\end{tabular}
\caption{\small Execution time (sec) of $k$-FSM:- $\sigma_{min}$: minimum support (OOM: out of memory; TO: timed out).}
\label{tab:fsm}
\end{table*}

\begin{figure*}[t]
	\centering
  \begin{minipage}{0.31\textwidth}
		\centering
		\includegraphics[width=0.99\textwidth]{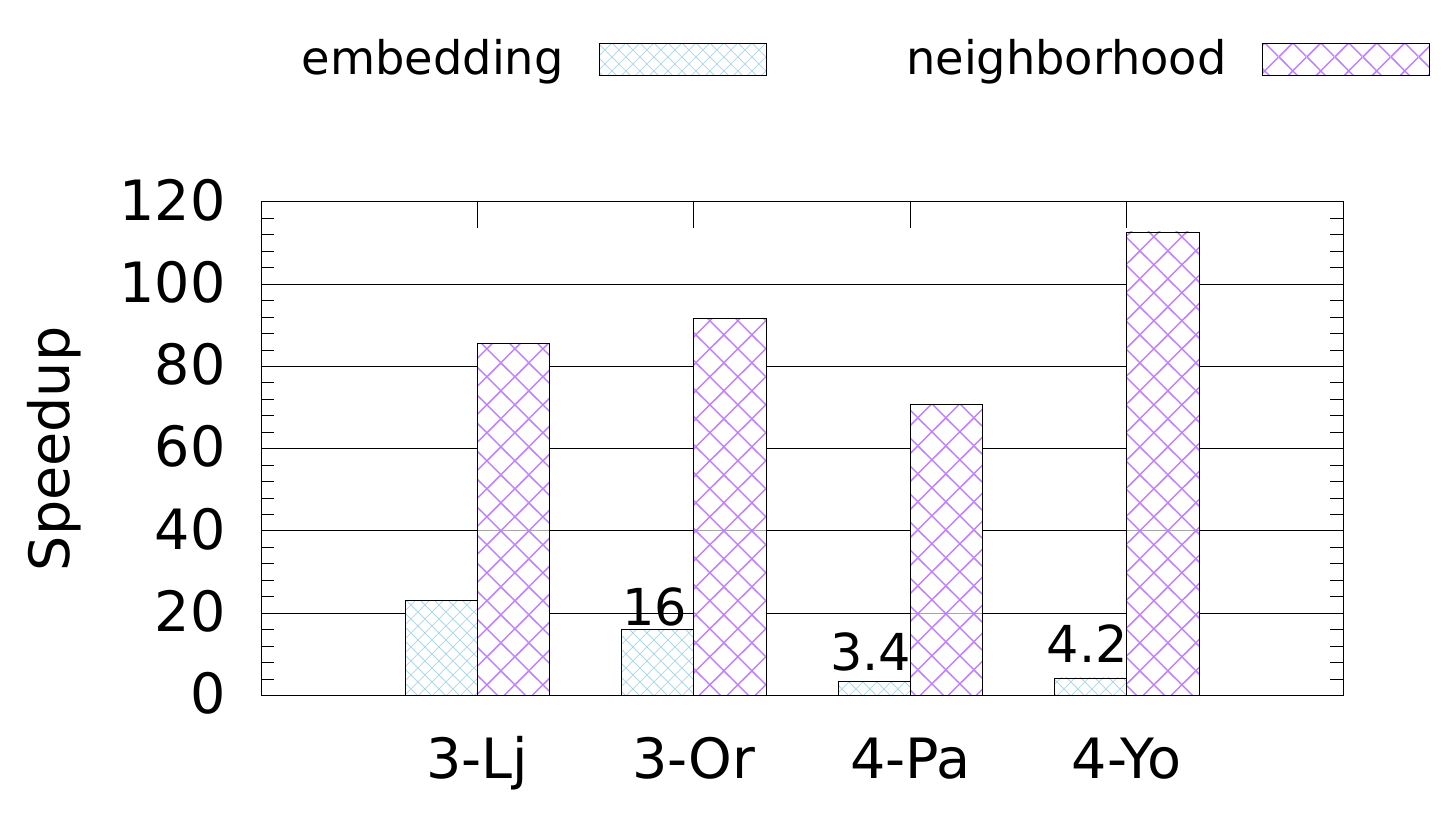}
		\caption{\small $k$-MC speedup with memoization of
			embedding/neighborhood connectivity.}
		\label{fig:mem-mc}
	\end{minipage}
  \hspace{3.6pt}
	\begin{minipage}{0.22\textwidth}
		\centering
		\includegraphics[width=\textwidth]{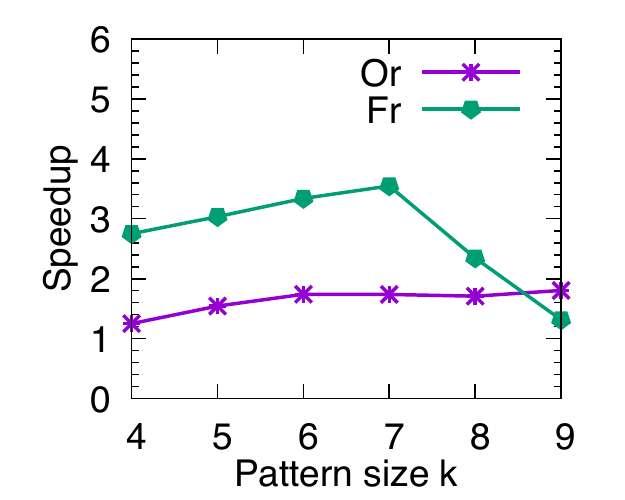}
		\caption{\small $k$-CL speedup by applying search on local graph.}
		\label{fig:shrink}
	\end{minipage}
\hspace{3.6pt}
	\begin{minipage}{0.37\textwidth}
		\includegraphics[width=\textwidth]{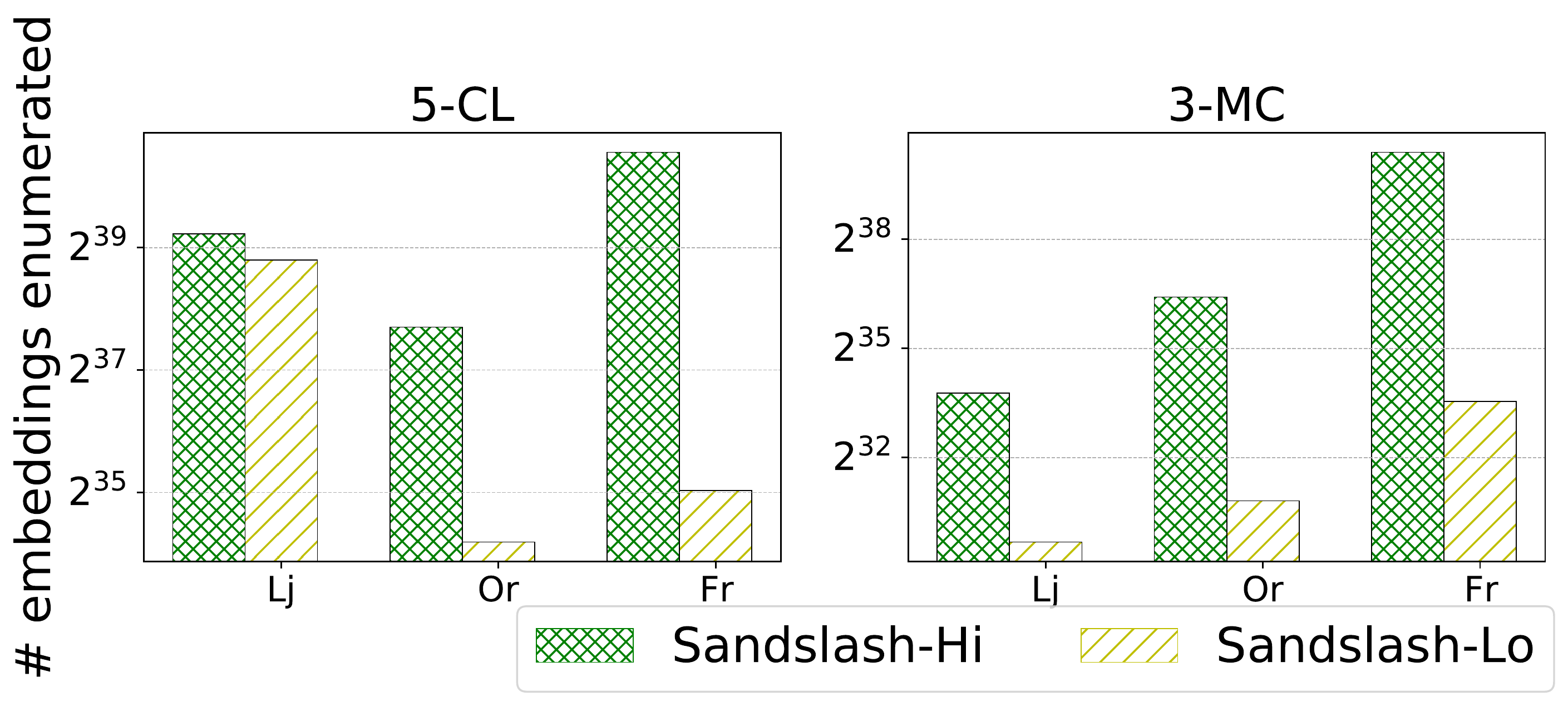}
		\caption{\small Comparing search space (\# of enumerated
			embeddings) of high- and low-level Sandslash.}
		\label{fig:space}
	\end{minipage}
\end{figure*}

\noindent{\textbf{$k$-Motif Counting ($k$-MC):}}
\cref{tab:mc} compares $k$-MC performance.
Sandslash-Hi outperforms AutoMine due to symmetry breaking, 
and Sandslash-Lo is orders of magnitude faster than Sandslash-Hi 
due to the local counting optimization. Pangolin is particularly inefficient 
for 4-MC as it cannot memoize neighborhood connectivity (MNC). 
Sandslash-Hi and Sandslash-Lo count all patterns simultaneously, 
whereas Peregrine does counting for each pattern/motif one by one, 
which allows it apply optimizations for each pattern. 
Unlike Sandslash, Peregrine reorders vertices during preprocessing.
Peregrine is faster than Sandslash-Hi likely due to these reasons. 
Sandslash-Lo is faster than Peregrine due to the formula-based local 
counting optimization, which cannot be supported in the Peregrine API.
All optimizations in expert-implemented PGD~\cite{PGD} are enabled in Sandslash-Lo.
Sandslash-Lo outperforms PGD because PGD does not apply 
symmetry breaking and has much larger enumeration space.
On average, Sandslash-Lo outperforms AutoMine, Pangolin, Peregrine, and
PGD by 27.2$\times$, 53.6$\times$, 8.6$\times$ and 17.9$\times$, respectively.



\noindent{\textbf{Subgraph Listing (SL):}}
\cref{tab:sl} presents SL results
(AutoMine is omitted since it has vertex-induced subgraph counting, not SL).
Sandslash outperforms all other systems, except 
Peregrine for the {\tt diamond} pattern on {\tt Lj} and {\tt Fr}, which is probably
because it reorders vertices during preprocessing. 
Pangolin on the other hand is much slower than the other systems as 
it does not support memoization of neighborhood 
connectivity (MNC) optimization.
The MNC approach in Sandslash 
is more efficient than the vertex set buffering (VSB) in  
Peregrine as explained in \cref{sec:agnostic-opt}:  
Peregrine must do neighborhood intersections to 
determine connectivity while Sandslash does not.
On average, Sandslash outperforms Pangolin and Peregrine by
29.5$\times$ and 5.6$\times$ respectively.



\noindent{\textbf{$k$-Frequent Subgraph Mining ($k$-FSM):}}
\cref{tab:fsm} presents $k$-FSM results
(AutoMine is omitted because it does not use domain support for FSM).
Although Peregrine uses DFS exploration, it does global 
synchronization among threads for each DFS iteration in FSM,
which essentially results in BFS-like exploration. 
In contrast, Sandslash uses DFS exploration on the sub-pattern tree
and filters patterns without synchronization.
Peregrine is the fastest for {\tt Yo} due to better 
load balance and relatively small number of frequent patterns. 
We observe that for graphs with a large number of frequent 
patterns ({\tt Pa}), Peregrine becomes very inefficient 
as its pattern-centric approach enumerates all the possible patterns 
first and then enumerates embeddings for each pattern one by one; 
this is detrimental to performance for larger graphs and patterns 
(e.g., it times out for {\tt Pdb}). Sandslash is similar or faster than 
Pangolin in most cases but is slower for {\tt Pa} at $\sigma$=30K 
mainly because the BFS based approach has high parallelism for that case.
For 4-FSM, Sandslash outperforms both Pangolin and Peregrine.
Sandslash performs better than expert-implemented DistGraph~\cite{DistGraph} too
as it automatically enables all optimizations that are in DistGraph.
Sandslash is the only system that can run 4-FSM on {\tt Pdb}.
On average, Sandslash outperforms Pangolin, Peregrine, and
DistGraph by 1.2$\times$, 4.6$\times$ and 2.4$\times$, respectively, for FSM.

\subsection{Analysis of Sandslash}\label{subsect:eval-opt}

Due to lack of space, we present only the impact of optimizations
in Sandslash that are missing in other systems (\cref{tab:sys_opt}).

\noindent{\textbf{High-Level Optimizations:}}
We observe 2\% to 16\% improvement for $k$-CL 
due to the degree filtering (DF) optimization.
\cref{fig:mem-mc} shows speedup due to
memoization of embedding connectivity (MEC) and
memoization of neighborhood connectivity (MNC) optimizations for $k$-MC.
For $k$-MC, the connectivity information in both 
the neighborhood and the embedding is memoized.
MEC and MNC improve performance by 7.4$\times$ and 87$\times$ on average.
\noindent{\textbf{Low-Level Optimizations:}}
Formula-based local counting (LC) reduces compute time by
avoiding unnecessary enumeration of patterns. \cref{tab:mc} shows
Sandslash-Lo is 38$\times$ faster than Sandslash-Hi due to LC. 
As the pattern gets larger, pruning becomes more important.
LC improves performance of 3-MC and 4-MC by 25$\times$ 
and 136$\times$ on average, respectively.
This highlights the need to expose a low-level interface
to express customized pruning strategies.

\cref{fig:shrink} illustrates the performance improvement on $k$-CL using
the local graphs (LG) optimization on large patterns. Shrinking the local graph
can reduce the search space compared to using the original graph.
This improves performance by 1.2$\times$ to 3.5$\times$ for {\tt Or} and
{\tt Fr}. The speedup for {\tt Or} increases as the
pattern size $k$ increases. However, for {\tt Fr}, the speedup
peaks at $k=7$, indicating that further shrinking becomes less
effective as $k$ grows. This trend depends on the input graph topology, but
in general, this optimization is effective for supporting large patterns.

Both LC and LG optimizations prune the enumeration search space.
We compare the search spaces of Sandslash-Hi and Sandslash-Lo to explain
how they improve performance.
\cref{fig:space} shows the number of enumerated embeddings for $k$-CL and
$k$-MC.  We observe a significant reduction for {\tt Or} and {\tt
Fr} in Sandslash-Lo, explaining the performance differences between
Sandslash-Hi and Sandslash-Lo in Tables~\ref{tab:cl} and~\ref{tab:mc}. However, the pruning is less
effective for {\tt Lj} in $k$-CL, and given the overhead of local graph
construction, Sandslash-Lo performs similar to Sandslash-Hi for {\tt
Lj} as shown in \cref{tab:cl}.

\begin{figure}[t]
  \begin{center}
    \includegraphics[width=0.48\textwidth]{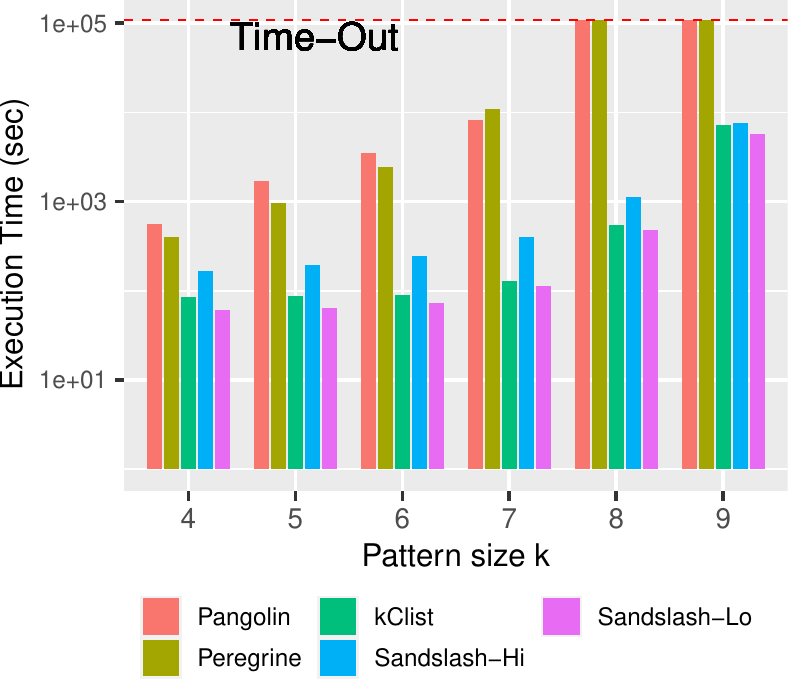}
    \caption{\small Execution time (sec in log scale) of $k$-CL on \texttt{Fr} graph.}
    \label{fig:large-kcl}
  \end{center}
\end{figure}


\noindent{\textbf{Large Patterns:}}
\cref{fig:large-kcl} shows $k$-CL on \texttt{Fr} graph with $k$ from 4 
to 9. 
Pangolin and Peregrine timed out for $k=8$ and $k=9$.
Existing systems cannot efficiently mine large patterns 
due to a much larger enumeration search space or 
significant amount of redundant computation. 
In contrast, Sandslash can effectively handle these large patterns,
and in all cases Sandslash-Lo is faster than expert-implemented kClist.

\noindent{\textbf{Large Inputs:}}
The large input graph, {\tt Gsh}, 
requires 199GB in Compressed Sparse Row (CSR) format on disk, 
so we evaluate it using 96 threads on the Optane machine. 
We were not able to run AutoMine and Peregrine on this large input. 
For 4-CL, Pangolin takes 6.5 hours, whereas Sandslash-Hi takes only 0.9 hours.
Sandslash-Hi's memory usage is low as well: peak memory usage for Sandslash-Hi 
is 436 GB, while Pangolin, a BFS-based system, uses 3.5 TB memory.
kClist and Sandslash-Lo run out of memory because 
maintaining the local graphs consumes more than 6 TB memory.

\noindent{\textbf{Strong Scaling:}}
The speedup of Sandslash on 56 threads over Sandslash on 1 thread is, 
on average, 43$\times$, 28$\times$, 39$\times$, 35$\times$, and 8$\times$ 
for TC, $k$-CL, SL, $k$-MC, and $k$-FSM, respectively. 
The speedup for $k$-FSM is lower due to constrained 
parallelism in traversing the sub-pattern tree. Sandslash balances 
work well because the number of grains/vertices is large enough. 
Orthogonal techniques like fine-grained work-stealing in Fractal and 
vertex reordering (preprocessing) in Peregrine can be  
added to Sandslash to further improve load balance.

\section{Related Work}\label{sect:relate}
\textbf{Low-level GPM Systems:}
Arabesque~\cite{Arabesque} is a distributed GPM system that proposed
the embedding-centric programming paradigm.
RStream~\cite{RStream} is an out-of-core GPM system on a single
machine. Its programming model is based on relational algebra.
Kaleido~\cite{Kaleido} is a single-machine system that
uses a compressed sparse embedding (CSE) data structure to
reduce memory consumption. G-Miner~\cite{G-Miner} is a distributed
GPM system which incorporates task-parallel processing.
Pangolin~\cite{Pangolin} is a shared-memory GPM system targeting
both CPU and GPU.
Instead of the BFS exploration used in the above systems,
Fractal~\cite{Fractal} uses DFS to enumerate
subgraphs on distributed platforms.
Compared to these systems, Sandslash improves
productivity and performance since many optimizations
are automated in Sandslash.
Some of these GPM systems use distributed, out-of-core,
or GPU platforms, which is orthogonal to our work.
The focus of our work is utilizing in-memory CPU platforms
to get the best performance.

\textbf{High-level GPM Systems:}
AutoMine~\cite{AutoMine} is a DFS based system targeting a single-machine.
It provides a high-level programming interface and employs
a compiler to generate high performance GPM programs.
Sandslash supports a wider range of GPM problems
and also enhances performance without compromising productivity.
Peregrine is the state-of-the-art high-level GPM system.
It includes efficient matching strategies from well-established
techniques~\cite{Cartesian,DUALSIM,Breaking} and
improves performance compared to previous systems.
Nevertheless, Sandslash outperforms Peregrine using only its high-level API.
Furthermore, Sandslash provides a low-level API to trade-off programming
effort for much better performance.


\textbf{GPM Algorithms:}
There are numerous hand-optimized GPM applications targeting
various platforms. For TC, there are parallel solvers on
multicore CPUs~\cite{Shun,GBBS,TC2017,TC2018}, distributed
CPUs~\cite{Suri,PDTL,TC2019}, and GPUs~\cite{TriCore,DistTC,H-INDEX}.
kClist~\cite{KClique} is a parallel $k$-CL algorithm derived from
\cite{Arboricity}. It constructs DAG using a core value based ordering
to reduce search space. PGD~\cite{PGD} counts 3 and 4-motifs by
leveraging proven formulas to reduce enumeration space.
Escape~\cite{ESCAPE} extends this approach to 5-motifs.
Subgraph listing~\cite{Ullmann,PSgL,CECI,DUALSIM,Cartesian,
PathSampling,TurboFlux,Ma,Lai,Scaling,FRDSE,EPSE,Graphflow,WCOJ,EmptyHeaded} is another
important application in which a matching order is
applied to reduce search space and avoid graph isomorphism tests.
gSpan~\cite{gSpan} is a sequential FSM algorithm using
a lexicographic order for symmetry breaking.
DistGraph~\cite{DistGraph,ParFSM}
parallelizes gSpan with a customized dynamic load
balancer that splits tasks on the fly.
We did holistic analysis on the optimizations introduced in these expert-written solvers
and implemented them in Sandslash for better productivity.


\section{Conclusion}\label{sect:concl}
In this work, 
We present Sandslash, 
a two-level shared-memory GPM system that provides high 
productivity and high performance without compromising expressiveness. 
The user can easily compose GPM applications with the 
support of automated optimizations and transparent parallelism.
The system also gives the user flexibility to express 
advanced optimizations to boost performance further.
Sandslash significantly 
outperforms existing systems and even hand-optimized implementations.

\bibliographystyle{plain}
\bibliography{references}

\clearpage
\newpage
\begin{appendices}

\section{Pseudo Code in Low-level Sandslash}\label{appendix:code}

\begin{lstlisting}[float=htp, floatplacement=tbp,
label={lst:motif-opt}, language=C++, abovecaptionskip=0pt,
caption={\small Sandslash-Lo user code for 3-MC using local counting.}]
void localReduce(int depth, vector<Support> &supports) {
	// accumulate local wedge count for each vertex
	if (depth == 0) {
		Vertex v = getHistory(depth);
		int n = getDegree(v);
		int pid = getWedgePid();
		supports[pid] += n * (n-1) / 2;
	}
}
Pattern p = generateTriangle();
Support tri_count = enumerate(p);
int pid = getTrianglePid();
supports[pid] = tri_count; // global triangle count
pid = getWedgePid();
supports[pid] -= 3 * tri_count; // global wedge count
\end{lstlisting}

\begin{lstlisting}[float=htp, floatplacement=tbp,
label={lst:4motif-lc}, language=C++, abovecaptionskip=0pt,
caption={\small Sandslash-Lo user code for 4-MC using local counting.}]
void localReduce(int depth, vector<Support> &supports) {
	// accumulate local counts for each edge
	if (depth == 1) {
		Vertex u = getHistory(0);
		Vertex v = getHistory(1);
		int deg_u = getDegree(u);
		int deg_v = getDegree(v);
		int tri = intersection_num(u, v);
		int staru = deg_u - tri - 1;
		int starv = deg_v - tri - 1;
		int pid = getDiamondPid();
		supports[pid] += tri * (tri - 1);
		pid = getTailedTrianglePid();
		supports[pid] += tri * (staru + starv);
		pid = get4PathPid();
		supports[pid] += staru * starv;
		pid = get3StarPid();
		supports[pid] += staru * (staru - 1) + starv * (starv - 1);
	}
}
Pattern p0 = generateClique(4);
Pattern p1 = generateCycle(4);
Set<Pattern> patterns({p0, p1});
Map<Pattern, Support> support_map = enumerate(patterns);
...
\end{lstlisting}


\begin{lstlisting}[float=htp, floatplacement=tbp,
label={lst:kcl-opt}, language=C++, abovecaptionskip=0pt,
caption={\small Sandslash-Lo user code for $k$-CL using the local graph.}]
// initialize the local graph by computing the 
// intersection of v's neighbors and u's neighbors, 
// where v is the start vertex, u is a neighbor of v's.
void initLG(Graph gg, Vertex v, Graph lg) { 
	for (u : gg.adjList(v)) 
		lg.insertVertex(u); // insert vertex u to the local graph
	// the system maintains the global to local ID mapping
	for (u : gg.adjList(v))
		for (w : (gg.adjList(u) & gg.adjList(v))) 
			lg.insertEdge(u, w); // insert edge (u, w) to local graph
}
// update the local graph by removing vertices not in
// the newly generated embedding list
void updateLG(int depth, Graph lg) {
	// for each vertex in the new embedding list
	for (u : this->emb_list[depth+1]) {
		// get the size of the old adjList
		int degree = lg.getDegree(depth, u);
		int idx = lg.edgeStart(u);
		int tail = eid + degree;
		// set the size of the new adjList to 0
		lg.setDegree(depth+1, u, 0);
		// traverse the neighbors of u
		for (w : lg.adjList(u)) {
			// if w is in the new embedding list
			if (this->emb_list[depth+1].contains(w)) {
				// w remains in the new adjList of u
				// size of u's new adjList increases by 1
				lg.incDegree(depth+1, u);
			} else {
				// remove w by swap it with
				// the vertex at the tail of the old adjList
				lg.insertAdj(idx--, lg.getEdgeDst(tail--);
				lg.insertAdj(tail, w);
			}
			idx ++;
		}
	}
}
\end{lstlisting}

\section{Details of Optimizations and Definitions}\label{appendix:opt}

\subsection{Symmetry Breaking (Partial Order)}\label{subsect:sym_break}
The problem of overcounting is circumvented by an {\em automorphism check}
which selects a {\em canonical} representation from identical subgraphs. 
This technique is also known as {\em symmetry breaking}.
A widely used approach is to apply partial orders between vertices 
in the embedding~\cite{DUALSIM,Peregrine}. 
For special patterns, e.g., cliques, symmetry breaking can be done 
by constructing a DAG, which avoids runtime checks but 
requires preprocessing. Sandslash supports both techniques 
and adaptively applies them according to the detected pattern.
In ~\cref{fig:subgraph-tree}, lightly colored subgraphs
are removed by automorphism checks, leaving a unique
canonical subgraph for each set of automorphisms.
Automorphism checks are also
useful for pruning the search tree,\emph{ e.g}., 
the subgraph (2,1) is not extended in \cref{fig:subgraph-tree}
because it is automorphic to the subgraph (1,2).
High-level Sandslash enumerates only canonical embeddings,
while low-level Sandslash allows user-defined automorphism check.

\subsection{Orientation (Total Order)}\label{subsect:dag}

The technique establishes
a total ordering among vertices, and converts the input graph into 
a Directed Acyclic Graph (DAG). The search is then done over this 
DAG instead of the original input graph, which substantially reduces 
the search space. In Sandslash, orientation is automatically enabled 
when Sandslash detects a clique as the input pattern, i.e., when 
$|E|=|V| \times (|V|-1)/2$ for a given pattern $P(V,E)$.
Currently two types of orientation schemes are supported: 
(1) degree based and (2) core value based. For degree based orientation, 
the ordering among vertices is established based on their degrees: 
each edge points towards the vertex with higher degree. 
When there is a tie, the edge points to the vertex with larger vertex ID.  
For core value based orientation, ordering is established using
the core value of vertices~\cite{KClique}, instead of degrees.
It saves memory for local graph search,
but requires more pre-processing time than the degree based orientation.

\begin{figure}[t]
	\begin{center}
		\includegraphics[width=0.43\textwidth]{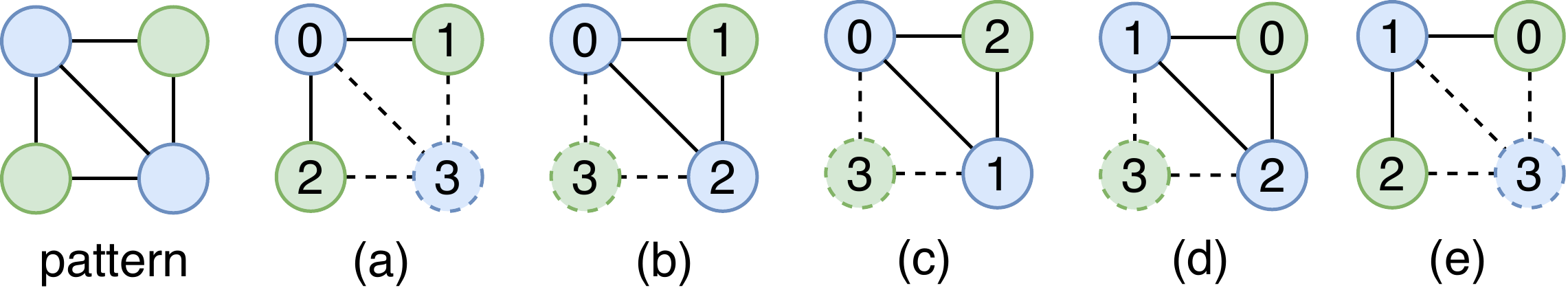}
		\caption{\small Possible matching orders for pattern
			 {\tt diamond}. The number in each vertex is not a vertex
			ID but the order being matched. Colors show the symmetric
			positions. The matching process can start from blue vertices
			(a \& b \& c) or green vertices (d \& e). Among them, (a) and (e)
			match a wedge first, and then form a diamond; (b) (c) and
			(d) discover a triangle first, and then form a diamond.}
		\label{fig:matching-order}
	\end{center}
\end{figure}

\subsection{Pattern-Guided Search (Matching Order)}\label{subsect:mo}

The pattern structure can be leveraged to prune the 
search space, which is known as \texttt{matching-order}.
\cref{fig:matching-order} shows the 5 possible 
matching orders for {\tt diamond}. We can find the triangle
first, and then add the fourth vertex connected to two of
the endpoints of the triangle. Another possible matching 
order is to find a wedge first, and then find the fourth 
vertex that is connected to all three vertices of the wedge. 
Both orders find the diamond, but they explore different parts 
of the search space. In real-world sparse graphs, the number of 
wedges is usually orders-of-magnitude larger than the 
number of triangles, so it is more efficient to find the 
triangle first. 


In general, using the right matching order reduces computation 
and memory consumption. Sandslash uses a greedy approach~\cite{DUALSIM} 
to choose a good matching order: at each step, (1) we choose a sub-pattern 
which has more internal partial orders; (2) If there is a tie, we choose a
denser sub-pattern, i.e., one with more edges. In \cref{fig:matching-order}, 
(c) is the matching order chosen by the system, since there is a partial order 
between vertex 0 and 1 for symmetry breaking (\ref{subsect:sym_break}). 
The intuition is that applying partial ordering as early as possible can better 
prune the search tree. Similarly, matching denser sub-pattern first can 
possibly prune more branches at early stage.

Note that using matching order avoids isomorphism test if the 
patterns are explicit. However, for implicit-pattern problems, 
{\em this approach needs to enumerate all the possible patterns before search}.
For example, FSM in AutoMine generates a matching order for each 
unlabeled pattern and includes a lookup table to distinguish between 
labeled patterns. This table is massive when the graph
has many distinct labels.
Peregrine also suffers significant overhead for FSM since there are 
many patterns and it matches each of them one by one.

\subsection{Fine-Grained Pruning}\label{subsect:fp}

Many GPM algorithms use their own pruning strategy during the
tree search, which significantly reduces the search space.
Sandslash exposes {\tt\small toAdd} and {\tt\small toExtend} 
to allow fine-grained control on the pruning strategies.
For example, in $k$-CL~\cite{KClique}, since an $i$-clique 
can only be extended from an ($i$-1)-clique,  
{\tt\small toExtend} and {\tt\small toAdd} can be used to 
only extend the last vertex in the embedding and check
if the newly added vertex is connected to all previous vertices in 
the embedding, respectively. In general, these two functions
are used to specify the matching order and partial orders
for explicit-pattern problems. The matching order is described
by {\tt\small toExtend} which specifies the next vertex to extend
in each level. Connectivity and partial orders are checked
in {\tt\small toAdd}. Sandslash generates these functions 
automatically for explicit-pattern problems.

\subsection{Customized Pattern Classification}\label{subsect:cp}
To recognize the pattern of a given embedding, a straightforward 
approach is the graph isomorphism test, which is an expensive computation. 
To avoid the test, Sandslash uses matching order for explicit patterns. 
For multiple explicit pattern problems, the connectivity check 
can be used to identify different patterns.
For small implicit patterns, Sandslash uses 
\emph{customized pattern classification} (CP) \cite{Pangolin}. 
For example, in FSM, the labeled {\tt \small wedge} patterns
can be differentiated by hashing the labels of the three vertices 
(the two endpoints of the wedge are symmetric).
To enable CP, the user needs to use the {\tt \small getPattern} 
API function to implement a customized method. 

\subsection{Example for Connectivity Code}\label{subsect:ccode}
As shown in \cref{fig:ccode}, if the bit at position $i$ of connectivity 
code is set, there is an edge in $G$ between vertex $u$ and 
the vertex contained in the ancestor at level $i$ in the tree. 

\begin{figure}[t]
\centering
\includegraphics[width=0.3\textwidth]{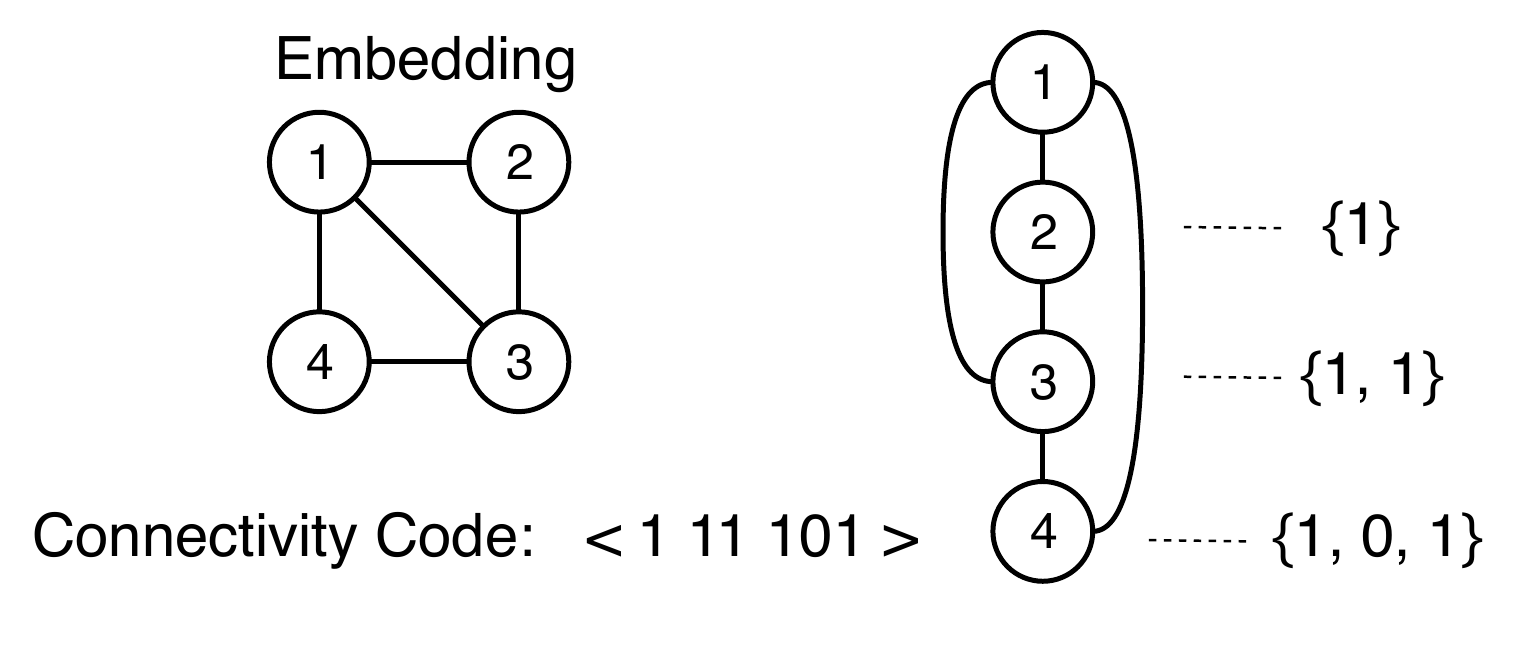}
\caption{\small Example of the connectivity code.
		The code of $v_2$ is \{1\}, indicating $v_2$ is
		connected with $v_1$. $v_4$ has a code of \{1,0,1\},
		meaning $v_4$ is connected to $v_1$ and $v_3$, but
		not $v_2$. Concatenating codes of all vertices, 
		we get the code of this embedding \{111101\}. 
		With this code we can rebuild
		the exact structure of the embedding on the left.}
\label{fig:ccode}
\end{figure}

\subsection{Subgraphs, Subgraph Trees and Patterns}\label{subsect:def}
The following definitions are standard~\cite{harary1969graph}.
Given a graph $G(V,E)$, a {\em subgraph} $G'(V',E')$ of $G$
is a graph where $V' \subseteq V$, $E' \subseteq E$. If $G_1$
is a subgraph of $G_2$, we denote it as $G_1 \subseteq G_2$.

Given a vertex set $W \subseteq V$, the {\em vertex-induced
subgraph} is the graph $G'$ whose (1) vertex set is $W$ and whose
(2) edge set contains the edges in $E$ whose endpoints are in $W$.
Given an edge set $F \subseteq E$, the {\em edge-induced
subgraph} is the graph $G'$ whose (1) edge set is $F$ and whose
(2) vertex set contains the endpoints in $V$ of the edges in $F$.
The difference between the two types of subgraphs can be
understood as follows. Suppose vertices $v_1$ and $v_2$
are connected by an edge $e$ in $G$. If $v_1$ and $v_2$ occur in
a vertex-induced subgraph, then $e$ occurs in the subgraph as
well; in an edge-induced subgraph, edge $e$ will be present only
if it is in the given edge set $F$. Any vertex-induced subgraph
can be formulated as an edge-induced subgraph.

Two graphs $G_1(V_1,E_1)$ and $G_2(V_2,E_2)$ are {\em isomorphic},
denoted $G_1 \simeq G_2$, if there exists a bijection $\mathbf{f}$:
$V_1 \rightarrow V_2$, such that any two vertices $u$ and $v$ of
$G_1$ are adjacent in $G_1$ if and only if $\mathbf{f}(u)$ and
$\mathbf{f}(v)$ are adjacent in $G_2$. In other words, $G_1$ and $G_2$
are structurally identical. In the case when $\mathbf{f}$ is a mapping
of a graph onto itself, i.e., when $G_1$ and $G_2$ are the
same graph, $G_1$ and $G_2$ are {\em automorphic}, i.e. $G_1 \cong G_2$.


\end{appendices}

\end{document}